\documentclass[%
 reprint,
 amsmath,amssymb,
 aps,
]{revtex4-1}

\usepackage{dcolumn}
\usepackage{bm}

\usepackage{amsmath, amssymb}
\usepackage{amsfonts}
\usepackage{braket}

\usepackage{ascmac}

\usepackage{graphicx}

\usepackage{color}

\newcommand{\eq}[1]{\begin{equation} #1 \end{equation}}
\newcommand{\eqa}[2]{\begin{equation} #1 \label{#2} \end{equation}}
\newcommand{\balign}[1]{\begin{align} #1 \end{align}}
\newcommand{\bcases}[1]{\begin{cases} #1 \end{cases}}

\newcommand{\figin}[4]
{\begin{figure}[t]
\centering
\includegraphics[width= #1]{#2.pdf}
\caption{#3}
\label{f:#4}
\end{figure}}

\newcommand{\todayd}{\the\year/\the\month/\the\day}
\newcommand{\del}{\partial}
\newcommand{\bib}{\bibitem}

\newcommand{\lmd}{\lambda}
\newcommand{\Lmd}{\Lambda}

\newcommand{\lb}{\label}
\newcommand{\nt}{\notag}
\newcommand{\Tr}{\mathrm{Tr}}

\newcommand{\bel}{\begin{easylist}}
\newcommand{\eel}{\end{easylist}}
\newcommand{\bi}[1]{\begin{itemize} #1 \end{itemize}}

\newcommand{\eref}[1]{Eq.~\eqref{#1}}

\newcommand{\fref}[1]{Fig.~\ref{f:#1}}
\newcommand{\sref}[1]{Sec.~\ref{s:#1}}

\def \({\left(}
\def \){\right)}
\newcommand{\la}{\left\langle}
\newcommand{\ra}{\right\rangle}
\newcommand{\abs}[1]{\left|#1\right|}

\newcommand{\sumtwo}[2]%
{\mathop{\sum_{#1}}_{#2}}
\newcommand{\sumthree}[3]%
{\mathop{\mathop{\sum_{#1}}_{#2}}_{#3}}
\newcommand{\sumfour}[4]%
{\mathop{\mathop{\mathop{\sum_{#1}}_{#2}}_{#3}}_{#4}} 
\newcommand{\prodtwo}[2]%
{\mathop{\prod_{#1}}_{#2}}
\newcommand{\mintwo}[2]%
{\mathop{\min_{#1}}_{#2}}
\newcommand{\maxtwo}[2]%
{\mathop{\max_{#1}}_{#2}}
\newcommand{\maxthree}[3]%
{\mathop{\mathop{\max_{#1}}_{#2}}_{#3}}
\newcommand{\limtwo}[2]%
{\mathop{\lim_{#1}}_{#2}}
\newcommand{\suptwo}[2]%
{\mathop{\sup_{#1}}_{#2}}
\newcommand{\supthree}[3]%
{\mathop{\mathop{\sup_{#1}}_{#2}}_{#3}}
\newcommand{\supfour}[4]%
{\mathop{\mathop{\mathop{\sup_{#1}}_{#2}}_{#3}}_{#4}} 
\newcommand{\inftwo}[2]%
{\mathop{\inf_{#1}}_{#2}}
\newcommand{\infthree}[3]%
{\mathop{\mathop{\inf_{#1}}_{#2}}_{#3}}
\newcommand{\inffour}[4]%
{\mathop{\mathop{\mathop{\inf_{#1}}_{#2}}_{#3}}_{#4}} 

\newcommand{\ep}{\varepsilon}

\newcommand{\Di}{\mathit{\Delta}}

\def\rnum#1{\resizebox{0.5em}{\height}{\expandafter{\romannumeral #1}}}
\def\Rnum#1{\resizebox{0.5em}{\height}{\uppercase\expandafter{\romannumeral #1}}}
\newcommand{\bH}{\beta^{\rm H}}
\newcommand{\bL}{\beta^{\rm L}}
\newcommand{\QH}{Q^{\rm H}}
\newcommand{\QL}{Q^{\rm L}}
\newcommand{\etac}{\eta_{\rm C}}
\newcommand{\htot}{H_{\rm tot}}
\newcommand{\hw}{H^{\rm W}}

\newcommand{\rhoc}[1]{\rho^{#1}_{\rm can}}
\newcommand{\rtot}{\rho_{\rm tot}}

\newcommand{\rtots}{\rho_{\rm tot'}}
\newcommand{\tle}{{\Lmd}_{\tilde{\rm E}}}
\newcommand{\lmdh}{\Lmd_{\rm H}}
\newcommand{\lmdl}{\Lmd_{\rm L}}
\newcommand{\lmde}{\Lmd_{\rm E}}

\newcommand{\rhoE}{\rho^{\rm E}}

\newcommand{\rhoi}{\rho_{\rm i}}
\newcommand{\rhof}{\rho_{\rm f}}

\newcommand{\hop}{H_{\rm op}}

\newcommand{\jH}{j_{\rm H}}
\newcommand{\jL}{j_{\rm L}}

\newcommand{\DS}{\Di S}
\newcommand{\etaC}{\eta_{\rm C}}

\newcommand{\rss}{\rho_{s(1)}}
\newcommand{\rst}{\rho_{s(t)}}
\newcommand{\vlr}{v_{\rm LR}}

\newcommand{\rhol}{\rho^{\rm L}}
\newcommand{\cdg}{c^{\dagger}}

\begin{document}

\preprint{APS/123-QED}

\title{Efficiency versus speed in quantum heat engines: Rigorous constraint from Lieb-Robinson bound}

\author{Naoto Shiraishi}
\affiliation{%
Department of Physics, Keio University, 3-14-1 Hiyoshi, Yokohama 223-8522, Japan
}%

\author{Hiroyasu Tajima}%
\affiliation{Center for Emergent Matter Science (CEMS), RIKEN, 2-1 Hirosawa, Wako, 351-0198 Japan}

\date{\today}

\begin{abstract}

A long standing open problem whether a heat engine with finite power achieves the Carnot efficiency is investigated.
We rigorously prove a general trade-off inequality on thermodynamic efficiency and time interval of a cyclic process with quantum heat engines.
In a first step, employing the Lieb-Robinson bound we establish an inequality on the change in a local observable caused by an operation far from support of the local observable.
This inequality provides a rigorous characterization of the following intuitive picture that most of the energy emitted from the engine to the cold bath remains near the engine when the cyclic process is finished.
Using the above description, we finally prove an upper bound on efficiency with the aid of quantum information geometry.
Our result generally excludes the possibility of a process with finite speed at the Carnot efficiency in quantum heat engines.
In particular, the obtained constraint covers engines evolving with non-Markovian dynamics, which almost all previous studies on this topics fail to address.

\begin{description}
\item[PACS numbers]
03.67.-a, 
05.30.-d, 
05.70.Ln, 
87.10.Ca,	
\end{description}
\end{abstract}

\pacs{Valid PACS appear here}
\maketitle

\section{Introduction}

Around 200 years ago, Carnot revealed that thermodynamic efficiency in a cyclic process with two heat baths with inverse temperatures $\bH$ and $\bL$ ($\bH<\bL$) is bounded by a universal function of these two temperatures, which was turned to be the celebrated Carnot efficiency \cite{Carnot, JT, callen} 
\eq{
{\eta}_{\rm C}=1-\frac{\bH}{\bL}.
}
At the same time, Carnot also showed that infinitely slow processes (i.e., zero power) realize the maximum efficiency.
An infinitely slow heat engine is however useless in practical sense, and thus a question naturally arises whether finite power engines attain the Carnot efficiency, which is the opposite statement of the aforementioned one.
Contrary to its apparent triviality, this has still been an open problem in spite of enormous effort for investigation of large power~\cite{Lyeo04, Casati07, Taylor15}, high efficiency~\cite{mahan, mahan-rev, majundar-rev, linke, dresselhaus, snyder, Casati08, auto, TH15}, and both of them~\cite{Sekimoto-Sasa, Aurell, Whitney, Raz16}.
We emphasize that although the question seems to be negative by intuition, it is hard to prove this rigorously in general setups.
In fact, conventional thermodynamics provides no restriction on the speed of processes, and even in the linear response regime the linear irreversible thermodynamics neither prohibits engines at the Carnot efficiency with finite power if time-reversal symmetry is broken~\cite{Benenti}. 
These findings elucidate the fact that the connection between heat exchange and dissipation is, surprisingly, not established even in the linear response regime when the framework of endoreversible thermodynamics~\cite{endo} no longer holds.

The latter work attracted renewed interest in this problem, and many researches are devoted to analyze specific models to find clues for capturing general properties on the relation between finite power and the Carnot efficiency.
One frequently-used setup to tackle the problem is a mesoscopic conducting system with non-interacting electrons under a magnetic field, where the incompatibility between finite power and the Carnot efficiency is shown through deriving a novel restriction on the Onsager matrix~\cite{Brandner, Brandner-full, Sothmann, Balachandran, Brandner-new, Yamamoto}.
Owing to the newly-derived restriction, it is shown both theoretically~\cite{Brandner, Yamamoto} and numerically~\cite{Brandner-new, Balachandran} that this system never attains the Carnot efficiency at finite power in the linear response regime.
Another frequently-used setup is a periodically driven system, for which the Onsager matrix can be defined and is in general asymmetric.
Utilizing the detailed information on microscopic dynamics, Refs.~\cite{underdamped, Bauer} and Refs.~\cite{Proesmans, Proesmans2} respectively demonstrate the incompatibility for underdamped Langevin particles with two heat baths and for isothermal driven systems.
On the other hand, some researches proposed ideas to realize these two simultaneously~\cite{Alla, EP, Campisi, Ponmurugan, PE} (some comments on these studies are seen in Ref.~\cite{remark}).
Recently, inspired by the idea of partial entropy production~\cite{SS, SIKS, SMS}, one of the authors has shown this incompatibility for general classical Markovian heat engines beyond the linear response regime~\cite{SST}.
However, non-Markovian heat engines have not been addressed in the previous works though real experimental heat engines are inevitably non-Markovian.
On the basis of the recent experimental development of small quantum heat engines~\cite{steeneken2010, crivellari2014, koski2014, rosnagel2015}, a general result covering non-Markovian heat engines is highly desired.

In this paper, we establish a trade-off relation between efficiency and speed of operation, which leads to this no-go theorem for general quantum heat engines.
We focus on the fact that most of the energy emitted from the engine to the cold heat bath remains in the region close to the engine, which leads to finite dissipation in the bath.
To prove this intuitive picture rigorously, we employ the Lieb-Robinson bound~\cite{LR, Has04, NOS06, HK, CSE08, Schuch11}, which claims that a commutator of two observables with different time acting on different regions far from each other cannot be large.
The Lieb-Robinson bound has recently used in various problems including problems on structure of a gapped ground state~\cite{HK, NS06, Has07}, the Lieb-Shultz-Mattis theorem in higher dimensions~\cite{Has04}, the quantum Hall effect~\cite{HM09}, and thermalization~\cite{IKS, KMS}.
Using this bound, we establish a useful upper bound on the change in a local observable caused by an operation with finite time-interval far from the area on which the local observable acts.
Applying this bound, we find that the expectation value of energy in the cold bath far from the engine does not largely change.
With help of the Pythagorean theorem for quantum relative entropy and some relations on the quantum Fisher information~\cite{AN, secondused}, we finally arrive at an upper bound for efficiency with the time interval of a cyclic process.
This bound clearly manifests that the faster the engine is operated the less the maximum efficiency becomes.

This paper is organized as follows.
In \sref{setup}, we show the setup of quantum heat engines and state our main inequality on efficiency and the time interval of a cyclic process.
Before going to the derivation, we sketch the outline of the proof in \sref{outline}.
The derivation is constructed with two part.
The former part is discussed in \sref{step1}, where we first introduce a useful lemma which bounds the effect of time-dependent operation, and using this we rigorously show the lower bound for the energy increase in a region near the engine.
The latter part is discussed in \sref{QIG}, where we connect the relative entropy to the energy increase, which leads to the desired inequality.
To examine how the obtained trade-off relation works, in \sref{ex} we demonstrate our inequality in a simple concrete example.
In \sref{transient}, we apply our analysis to a transient process and derive an extended version of the principle of maximum work.

We note that the obtained inequality turns to be no more than the second law of thermodynamics in the Markovian limit, in which the Lieb-Robinson velocity diverges.
Therefore, we treat Markovian heat engines in a completely different way.
We extend the result of classical Markov processes shown in Ref.~\cite{SST} to quantum Markov processes, which denies the compatibility between finite power and the Carnot efficiency.
In Appendix.\ref{s:markov}, we derive an inequality on efficiency and power for quantum Markovian heat engines.

\section{Setup and main result}\lb{s:setup}

\figin{7cm}{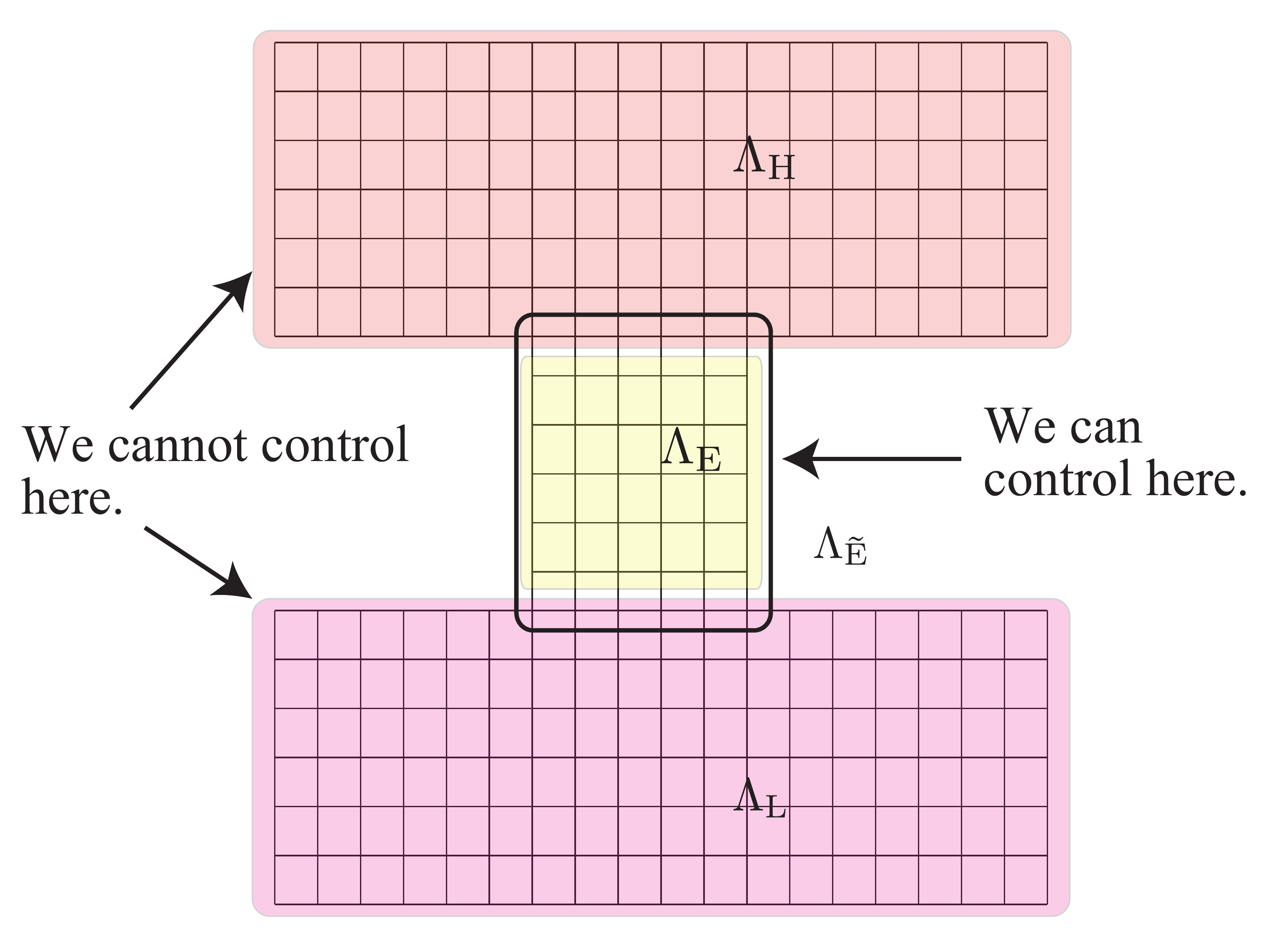}{
Schematic of the total system.
Three subsets of lattices $\lmde$, $\lmdh$, $\lmdl$ correspond to the regions of the engine (yellow), the hot bath (orange), and the cold bath (pink).
The subset $\tle$ is a composite set of $\lmde$ and its nearest-neighbor sites.
A time-dependent Hamiltonian acts only on $\lmde$.
}{f:setup}

We first describe our setup of heat engines.
In the most part of this paper, we restrict our attention to a quantum lattice system.
A lattice system is known as a manageable stage to investigate rigorous thermodynamic properties of nonequilibrium dynamics, and in most cases lattice systems keep universal thermodynamic properties.
This is why many rigorous results on thermodynamic properties of nonequilibrium dynamics are studied in such systems~\cite{IKS, KMS, CC, CDEO, ISPU, MAMW}.
We remark that recent sophisticated techniques indeed realize such lattice systems experimentally~\cite{BDZ, Tro, Kau}.

We in particular consider a composite system of an engine E and two heat baths H and L with inverse temperatures $\bH$ and $\bL$ ($\bH<\bL)$ (see \fref{f:setup}).
Let $\lmde$, $\lmdh$, $\lmdl$ be subsets of the sites in a region corresponding to the engine, the hot bath, and the cold bath, respectively.
$\lmde$ is attached to both $\lmdh$ and $\lmdl$, and $\lmdh$ is not attached to $\lmdl$.
We also define $\tle$ as a set of sites in $\lmde$ and its nearest-neighbor ones.
For convenience, we assume that the size of the engine $|\tle|$ is finite.
The treatment of thermodynamic limit is briefly discussed in \sref{dis}.

We now describe the dynamics of the system.
We suppose that baths are not driven, and interact only with the engine.
This is because our interest is on engines satisfying conventional thermodynamics, where a bath itself is an isolated system, not externally-driven~\cite{comment-driven}.
If a bath is driven, the system may violate the zeroth law and the second law of thermodynamics~\cite{KSH, DBS}.
Therefore, we do not consider externally-driven heat baths, and safely assume that the Hamiltonian of the composite system $\htot (t)$ is short-range interaction.
For simplicity, we consider the case of the sum of one-body Hamiltonians and nearest-neighbor two-body interaction Hamiltonians in the main part.
The case with general Hamiltonians with short-range interaction is analyzed in Appendix.\ref{s:gen}.
In a cyclic process in $0\leq t\leq \tau$, the Hamiltonian is changed with time as satisfying $\htot (0)=\htot (\tau)$.
Suppose that the Hamiltonian is time-dependent only on $\tle$.
We then decompose the total Hamiltonian $\htot (t)$ into five parts as 
\eq{
\htot (t)=H^{\rm E}(t)+H^{\rm EH}(t)+H^{\rm EL}(t)+H^{\rm H}+H^{\rm L},
}
where $H^X$ ($X=$ E, H, L) acts only on $X$, and $H^{{\rm E}X}$ ($X=$ H, L) is the sum of all interaction Hamiltonians between a site in $\lmde$ and that in $\Lmd_X$.
The Hamiltonians of the baths are set to be time-independent because we consider the situation that the external operation acts only on the engine and the bath is not driven as explained above.
We also suppose that the engine is initially separated from baths: $H^{\rm EH}(0)=H^{\rm EL}(0)=0$, which is a widely-used setup in the context of nonequilibrium statistical mechanics~\cite{TH15, QFT, CHT, Brandao}.
This setup corresponds to the situation that the initial state of the engine and that of the baths (i.e., canonical distributions) are prepared independently.
We do not impose any restriction on the Hamiltonian except the aforementioned ones.

We denote the density matrix of the total system at time $t$ by $\rtot (t)$, and write $\rho_{\rm i}:=\rtot (0)$ and $\rho_{\rm f}:=\rtot (\tau)$.
The partial trace of $\rho$ to $X$ ($X=$ E, H, L) is denoted by $\rho^X$.
Using the canonical distribution of baths $X$ expressed as $\rhoc{X}:=e^{-\beta^XH^X}/Z^X$ ($X=$H,L) with $Z^X:=\Tr[e^{-\beta^XH^X}]$, we set the initial state as a product state $\rho_{\rm i}=\rhoE_{\rm i}\otimes \rhoc{\rm H}\otimes\rhoc{\rm L}$, where $\rhoE_{\rm i}$ is arbitrary.
The cyclicity of the process requires that the final and the initial state of the engine have the same energy expectation value $\Tr[ H^{\rm E}(\tau)\rhoE_{\rm f}]=\Tr[ H^{\rm E}(0)\rhoE_{\rm f}]=\Tr[ H^{\rm E}(0)\rhoE_{\rm i}]$ and the same von Neumann entropy $S(\rhoE_{\rm f})=S(\rhoE_{\rm i})$~\cite{req-cycle}. 
We remark that if the above conditions are satisfied, the initial and final density matrices can be different.
The law of energy conservation tells that the extracted work $W$ is given by
\eqa{
W:=\Tr[\htot(0)\rtot (0)]-\Tr[\htot(\tau)\rtot(\tau)].
}{def-work}
The heat released from the bath H and that absorbed by the bath L are respectively written as
\balign{
\QH&:=\Tr[H^{\rm H}(\rho^{\rm H}_{\rm i}-\rho^{\rm H}_{\rm f})], \\ 
\QL&:=\Tr[H^{\rm L}(\rho^{\rm L}_{\rm f}-\rho^{\rm L}_{\rm i})].
}
Since there is no interaction between the engine and the baths in the initial and final states, the above definitions of work and heat contain no ambiguity.
We note that although our setup and result are described in terms of externally-operated heat engines, the same inequality holds for an autonomous evolution with a work storage, a catalyst, and a clock, which is frequently discussed in the field of quantum thermodynamics~\cite{TH15, Brandao, HO, SSP, TWO, oneshot1, oneshot2, oneshot3, Car2, Popescu2015, review, catalyst, Malabarba, m-based, T-W, MTH, Tasaki2015} (details are demonstrated in Appendix.~\ref{storage}).

We now state our main result.
In the above setup, the efficiency $\eta:=W/\QH$ is bounded by
\eqa{
\eta \leq {\etaC}-\frac{b}{( 2v_{\rm LR}\tau +C)^D}
}{mainnm}
with
\eq{
b:=\frac{(Q^{\rm L})^2}{8\beta^{L}\QH j_{\rm max}}.
}
Here, $D$ is the spatial dimension, and $j_{\rm max}$, $v_{\rm LR}$ are positive constants.
Roughly speaking, $j_{\rm max}$ corresponds to the heat capacity of the bath per unit volume and $v_{\rm LR}$ corresponds to the maximum speed of information propagation in the bath L, which is sometimes called a counterpart of the speed of light in non-relativistic quantum theory.
$C$ is a correction term depending on the details of the bath and the engine, which is irrelevant for the case of large $\tau$.
In this case, the obtained bound is roughly regarded as 
\eqa{
\eta \leq {\etaC}-\frac{b}{( 2v_{\rm LR}\tau)^D}.
}{clearer-main}
The precise definition of these constants are given in \sref{step1}, and a concrete example is demonstrated in \sref{ex}.
The inequality \eqref{mainnm} (or \eqref{clearer-main}) exhibits the existence of a trade-off between speed and efficiency, and in particular shows that a finite power heat engine (i.e., $\tau<\infty$) never attains the Carnot efficiency as long as the coefficients are finite.

We finally confirm the finiteness of coefficients.
First, the Lieb-Robinson velocity is finite ($\vlr <+\infty$) if the operator norm of local Hamiltonians of the bath is finite.
The case of Markovian limit, where $\vlr$ diverges, is considered in Appendix.~\ref{s:markov} with a completely different approach.
Second, $j_{\rm max}$ is finite if the heat capacity of the bath is finite.
Thus, if the bath is not at the first-order phase transition point, $j_{\rm max}$ is finite.
Other quantities ($\QL, \QH, \bL$) are obviously finite.
We thus conclude that if the operator norm of the Hamiltonian of the bath is finite and the baths are not at the first-order phase transition point, Eqs.~\eqref{mainnm} and \eqref{clearer-main} are indeed stronger than the Carnot bound.
Although, the difference from the Carnot bound might be small, it is indeed finite and the approach to the Carnot efficiency is fundamentally prohibited.

\figin{8cm}{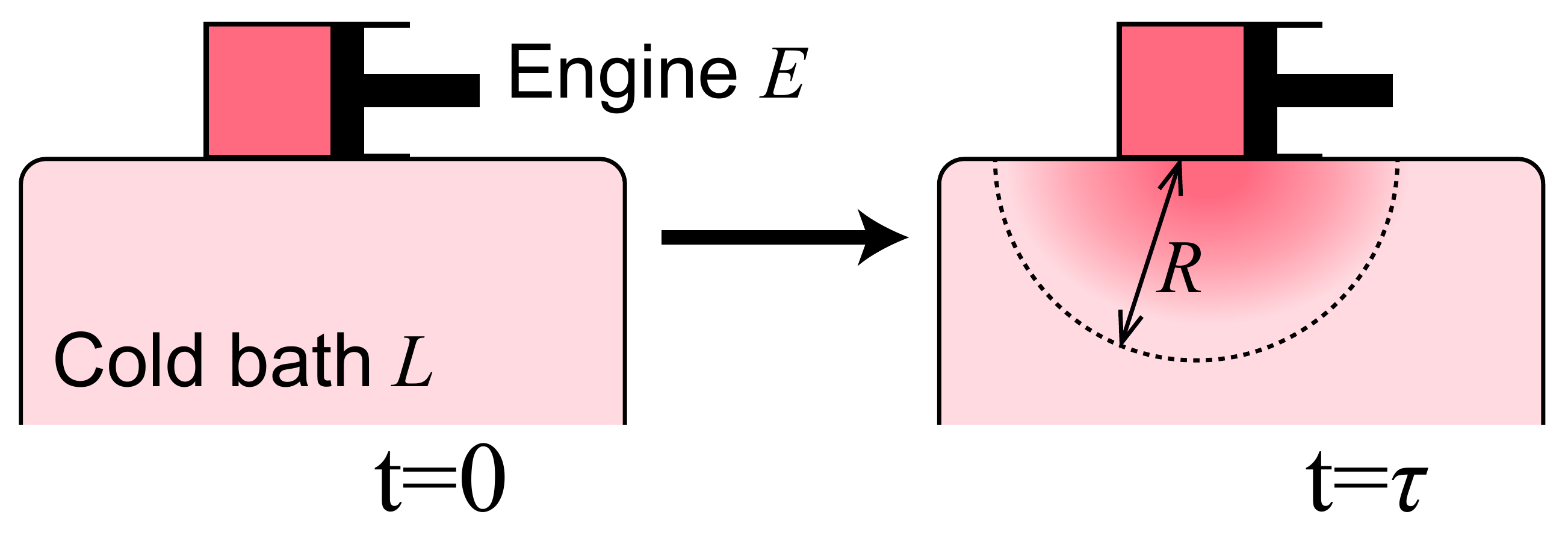}{
The key fact to derive the bound on efficiency.
Most of the heat emitted from the engine to the cold bath does not escape from the region close to the engine by the end of the operation $t=\tau$.
}{Power}

\section{Outline of the proof}\lb{s:outline}

Our starting point to prove \eqref{mainnm} is an expression of efficiency with quantum relative entropy~\cite{NC}.
Since the von Neumann entropy $S(\rho ):=-\Tr[\rho \ln \rho]$ is invariant under unitary evolution, we rewrite the quantum relative entropy between $\rho_{\rm f}$ and $\rho_{\rm i}$ as 
\eq{
D(\rho _{\rm f}||\rho _{\rm i}):=\Tr [\rho _{\rm f}(\ln \rho _{\rm f}-\ln \rho_{\rm i})]=\Tr [(\rho _{\rm i}-\rho _{\rm f})\ln \rho _{\rm i}].
}
A simple calculation yields the following inequality:
\balign{
&\Tr [(\rho _{\rm i}-\rho _{\rm f})\ln \rho _{\rm i}] \nt \\
=&-\bH \Tr [(\rho _{\rm i}^{\rm H}-\rho _{\rm f}^{\rm H})H^{\rm H}]-\beta ^{\rm L}\Tr [(\rho _{\rm i}^L-\rho _{\rm f}^L)H^{\rm L}] \nt \\
&-D(\rhoE_{\rm f}||\rhoE_{\rm i}) \nt \\
\leq&-\bH \QH+\bL\QL,
}
where we used the cyclic property $S(\rhoE_{\rm f})=S(\rhoE_{\rm i})$ and nonnegativity of relative entropy. 
Combining these two relations, we obtain a useful relation on efficiency:
\balign{
\eta \leq&1-\frac{\bH}{\bL}-\frac{D(\rho _{\rm f}||\rho _{\rm i})}{\bL \QH}. \lb{eta}
}
We remark that the quantum fluctuation theorem~\cite{QFT} leads to the expression of entropy production with relative entropy, which also implies the above relation.
Our problem is thus how to evaluate the above relative entropy in terms of microscopic description of the system.

The evaluation is accomplished in two steps.
In the first step, we use the fact that most of the emitted energy to the bath L remains in the region L close to the engine E (see \fref{Power}).
To state it rigorously, let L$'$ be a region in L whose distance from the engine $\tle$ is less than $R$.
Then, the precise statement is that the energy increase in L$'$ is bounded below by $\QL/2$ with $R\geq R^*$, where $R^*$ is given in an explicit form.
The quantity $( 2v_{\rm LR}\tau +C)^D$ in \eref{mainnm} corresponds to the upper bound of the volume of L$'$.

In the second step, we calculate the relative entropy $D(\rho _{\rm f}||\rho _{\rm i})$.
We here provide an intuitive idea of calculation instead of a rigorous proof, which is shown in \sref{QIG}.
First, using the monotonicity of the relative entropy~\cite{NC}, we have $D(\rho _{\rm f}||\rho _{\rm i})\gtrsim D(\rho _{\rm f}^{{\rm L}'}||\rho _{\rm i}^{{\rm L}'})$.
Second, we regard that the initial state of L$'$ is the canonical distribution: $\rho _{\rm i}^{{\rm L}'}\simeq \rho _{\rm can}^{{\rm L}'}$.

We now recall a property of relative entropy with a canonical distribution.
Let $\rho_{\rm can}$ be a canonical distribution with inverse temperature $\beta$ and $\rho'$ be an arbitrary density matrix, whose expectation values of energy are respectively denoted by $E$ and $E'$.
Then, the relative entropy is evaluated as
\eq{
D(\rho'||\rho_{\rm can})\geq D(\rho'_{\rm can}||\rho_{\rm can})\geq\frac{(E-E')^2}{2C_{\rm max}},
}
where $\rho'_{\rm can}$ is a canonical distribution with inverse temperature $\beta'$ whose expectation value of energy is $E'$, and $C_{\rm max}$ is the maximum heat capacity of the system with inverse temperature between $\beta$ and $\beta'$.
In the first inequality, we used the Jaynes's maximum entropy principle~\cite{Jay} that under a given energy expectation value $E'$ the von Neumann entropy is maximized for the canonical distribution.

The presented explanation is not a proof because the aforementioned approximation does not hold.
In \sref{QIG}, we derive the lower bound of relative entropy by employing the Fisher information.
Roughly speaking, the Fisher information corresponds to the heat capacity.
$j_{\rm max}$ in \eref{mainnm} is the maximum of the Fisher information normalized by the volume.


\section{Step 1: Bound on energy change of bath in the vicinity of engine}\lb{s:step1}

\subsection{Lieb-Robinson bound}\lb{s:LR}

\figin{8cm}{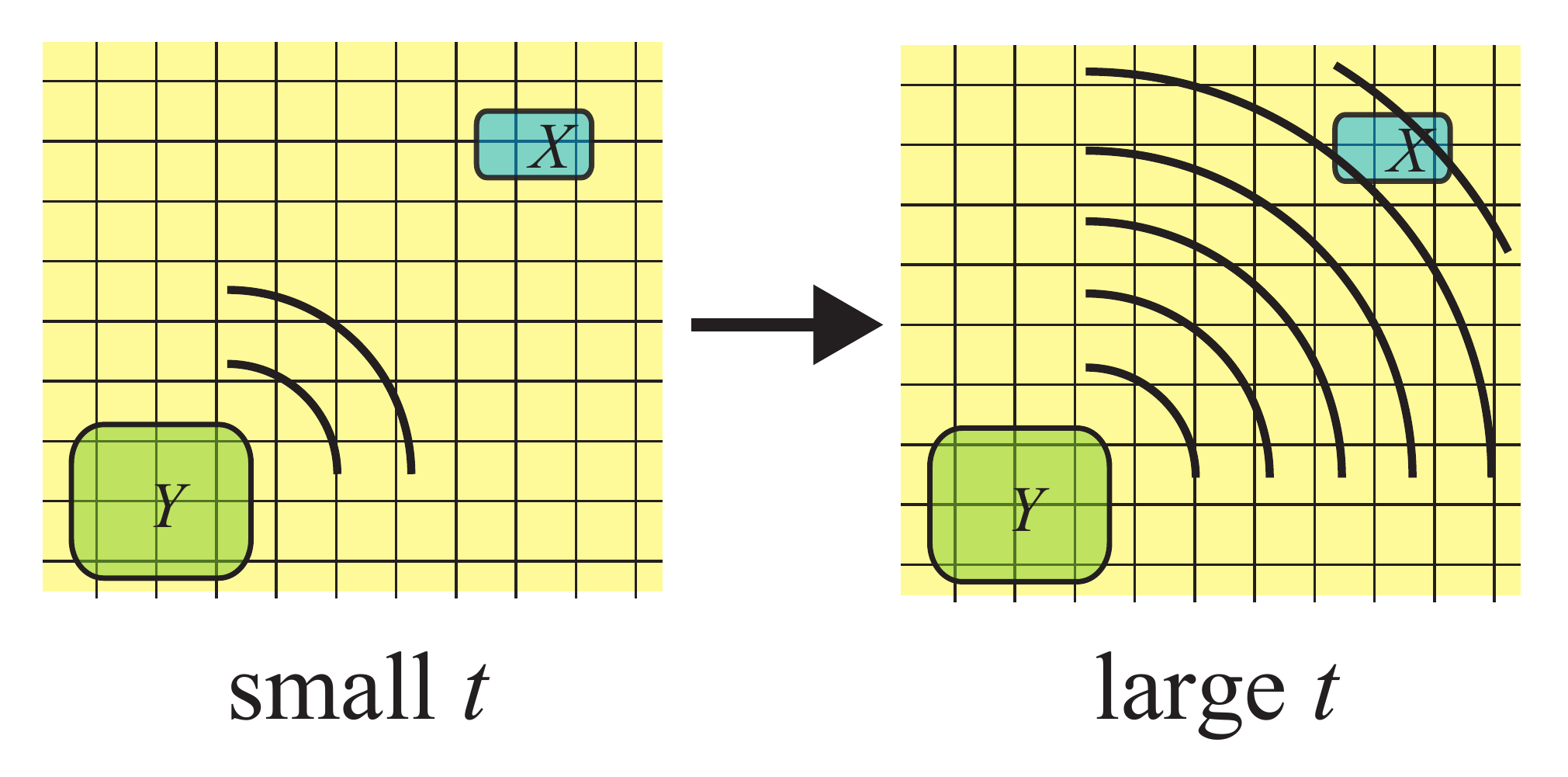}{
Schematics of the Lieb-Robinson bound.
The effect in the region $Y$ cannot be detected in the region $X$ far from $Y$ unless we wait sufficiently long time.
}{f:LR}

Before going to the proof of the bound, we explain the celebrated Lieb-Robinson bound for spin or fermionic systems~\cite{LR, HK}.
(The case of bosonic systems is investigated in Refs.~\cite{CSE08, Schuch11}.)
The Lieb-Robinson bound provides a restriction on the commutator between two operators acting on different space-time points (A schematic is seen in \fref{f:LR}).
This bound implies that two distant points with short time difference have almost zero correlation.
This is why some people say the Lieb-Robinson bound as the non-relativistic counterpart of causality in the relativistic theory.

We here introduce some quantities related to the lattice.
As following Ref.~\cite{HK}, we suppose that every bonds have their length 1 for simplicity.
We then define a distance between two sites $x$ and $y$ denoted by $d(x,y)$ as the length of the minimum path from $x$ to $y$.
We next introduce the sphere with its center $x$ and its radius $r$ as
\eq{
S(r;x):=\{ y|d(x,y)=r\} .
}
Using this, we define the dimension of the lattice $D$ as the minimum number satisfying the following condition for all $x$ and $r$ with a constant $K$:
\eqa{
|S(r;x)|\leq Kr^{D-1}.
}{def-dim}

Suppose a quantum spin or fermionic system on a discrete lattice.
Its Hamiltonian is written as the sum of the time-independent local Hamiltonians: $H =\sum_Z h_Z$, where $Z$ runs all finite subsets of sites and $h_Z$ is a local Hamiltonian whose support is $Z$.
We assume that for any $x$ and $y$
\eqa{
\sum_{Z\ni x,y}\| h_Z\| \leq \lmd e^{-\mu d(x,y)}
}{expdecay}
is satisfied with constants $\lmd$ and $\mu$, where $\| \|$ represents the operator norm~\cite{def-norm}.
The distance between two regions $X$ and $Y$ is defined as $d(X,Y):=\min_{x\in X, y\in Y}d(x,y)$.
The Lieb-Robinson bound claims that any bounded operators on $X$ and $Y$ denoted by $A_X$ and $B_Y$ satisfy
\eqa{
\| [A_X(t), B_Y]\| \leq c\| A_X\| \|B_Y\| |X||Y| e^{-\mu d(X,Y)} \( e^{vt}-1\)
}{LR}
with finite constants $c$ and $v$.
The coefficient $c$ depends only on the Hamiltonian of the system and the metric of the lattice~\cite{def-c}, and $v$ is given by $v:=4\lmd /c\hbar$.
$A_X(t)$ is the Heisenberg picture of $A_X$ at time $t$.
We define the {\it Lieb-Robinson velocity} as
\eq{
\vlr :=\frac{v}{\mu},
}
which characterizes the maximum speed of the information propagation in this system.
The Lieb-Robinson velocity is the non-relativistic counterpart of the speed of light.

\subsection{Bound on change in a local operator far from the time-dependent region}\lb{s:info-prop}

We now go back to the original problem.
Let us decompose the total Hamiltonian $\htot (t)$ into the time-independent part $H_0:=\htot (0)$ and the time-dependent part $H_{\rm op}(t):=\htot (t)-H_0$.
We remark that $\hop(t)$ acts on only $\tle$, and $H_0$ keeps the  initial state of the bath L, $\rhoc{\rm L}$, invariant.
Thus, as long as we are interested in quantities of the bath L, we can interpret the problem of the comparison of density matrices at two different times, $t=0$ and $t=\tau$ (i.e., $\rho_{\rm i}$ and $\rho_{\rm f}$), into the comparison between those with two different operations, $H_0$ and $H_0+H_{\rm op}(t)$, at the equal-time $t=\tau$.
To solve this problem, we here introduce a useful lemma which bounds the amount of information propagation from an operation with finite time interval.

\

{\bf Lemma}:
Consider a lattice system with single-body and two-body nearest-neighbor interaction Hamiltonians.
The full Hamiltonian $\htot (t)$ is supposed to be decomposed as $\htot (t)=H_0+H_{\rm op}(t)$, where $H_0$ is time-independent and $H_{\rm op}(t)$ acts only on a finite region $Y$.
Let $X$ be a finite region which does not have overlap with $Y$, and $A_X$ be a bounded operator on $X$.
We here consider two different time evolutions: 
In both cases, the initial state is set to the same state $\rho_{\rm i}$.
In the first case, the state evolves to $t=\tau$ under the Hamiltonian $\htot (t)$, while in the second case it evolves under $H_0$.
The density matrices under these two time evolutions are denoted by $\rho^{\rm tot}(t)$ and $\rho^0(t)$, respectively.
Then, the difference of the expectation value of $A_X$ between $\rho^0$ and $\rho^{\rm tot}$ at $t=\tau$ is bounded as
\balign{
&|\Tr [A_X\rho ^0(\tau )]-\Tr[A_X\rho ^{\rm tot}(\tau )]| \nt \\
\leq& \frac{c}{\hbar}H_{\rm op}^{\rm max} \| A_X\| |Y||X|e^{-\mu d(Y,X)} \( \frac{e^{v\tau}-1}{v}-\tau \) , \lb{info-prop}
}
where $H_{\rm op}^{\rm max} :=\max_t \| H_{\rm op}(t)\|$ and $c$, $\mu$, $v$ are the same constants as those in the Lieb-Robinson bound.

\

We emphasize that this relation \eqref{info-prop} is not obtained by a naive application of the Lieb-Robinson bound.
This is because the Lieb-Robinson bound connects observables on two space-time points while \eref{info-prop} treats the effect of an operation with finite time interval.
To obtain an inequality for such a case, we need to complement carefully $N-1$ new Hamiltonians between $H_0$ and $H_0+H_{\rm op}(t)$.
The lemma is proven in the Appendix.~\ref{Lemma1}.

We then rigorously show that most of the energy emitted to the bath L remains near the engine.
We fix a site $y$ in the engine and write $l:=\max_{x\in \tle}d(y,x)$ as the maximum distance to any site in $\tle$ from $y$.
Then, a site $x$ in L with $d(x,y)=R$ is at least $R-l$ far from the engine.
We now divide the bath L into two regions: 
\bi{
\item Region 1: sites $x$ with $d(x,y)\leq R$
\item Region 2: sites $x$ with $R+1 \leq d(x,y)$.
}
The region 1 corresponds to the region L$'$ in \sref{outline}.
The sets of lattices in the region 1 and 2 are labeled as $\Lmd^1$ and $\Lmd ^2$, respectively.
Correspondingly, we decompose $H^{\rm L}$, the Hamiltonian on L, into three parts, $H^1$, $H^2$, and $H^{12}$, which are the sum of local Hamiltonians with its support on only $\Lmd^1$, $\Lmd^2$, and both $\Lmd^1$ and $\Lmd^2$, respectively.
The total energy in L, $\Tr[H^{\rm L}\rho^{\rm L}]$, is decomposed into three parts: $E_1:=\Tr [H^1\rho^{\rm L}]$, $E_2:=\Tr [H^2\rho^{\rm L}]$, and $E_{12}:=\Tr [H^{12}\rho^{\rm L}]$.
We denote the difference of energy $E_a$ ($a=1,2,12$) between $\rhoi$ and $\rhof$ by $\Di E_a$.

We set $R$ larger than $(D-1)/\mu+1$.
Summing up \eref{info-prop} with respect to $A_X$ for all local Hamiltonians related to sites in the region 2, we find the following result (Calculation is shown in Appendix.\ref{derive-DiE}):
By setting $R$ as
\eqa{
R\geq R^*:=2\vlr \tau +C,
}{Rineq}
with a constant
\eq{
C=\frac{2}{\mu}\ln \frac{KcH_{\rm op}^{\rm max} H_{\rm site}^{\rm max}2|\tle| 2^{D}(D-1)!}{v\hbar\mu^D\QL}+2l+1,
}
we find $\Di E_{12}+\Di E_2\leq \QL/2$, which implies that at least energy of $\QL/2$ remains in the region 1: 
\eqa{
\Di E_1>\QL/2.
}{DiE1}
Here, we defined 
\eq{
H_{\rm site}^{\rm max}:=\sum_y\sum_{Z\ni x,y}\| h_Z\| 
}
as the upper bound of the summation of all the energy related to a single site (i.e., one-body Hamiltonian on the site and two-body Hamiltonians acting on this site and another site).
The constants $K$ and $D$ are given in \eref{def-dim}.

\section{Step 2: Energy change and quantum relative entropy}\lb{s:QIG}

\figin{7cm}{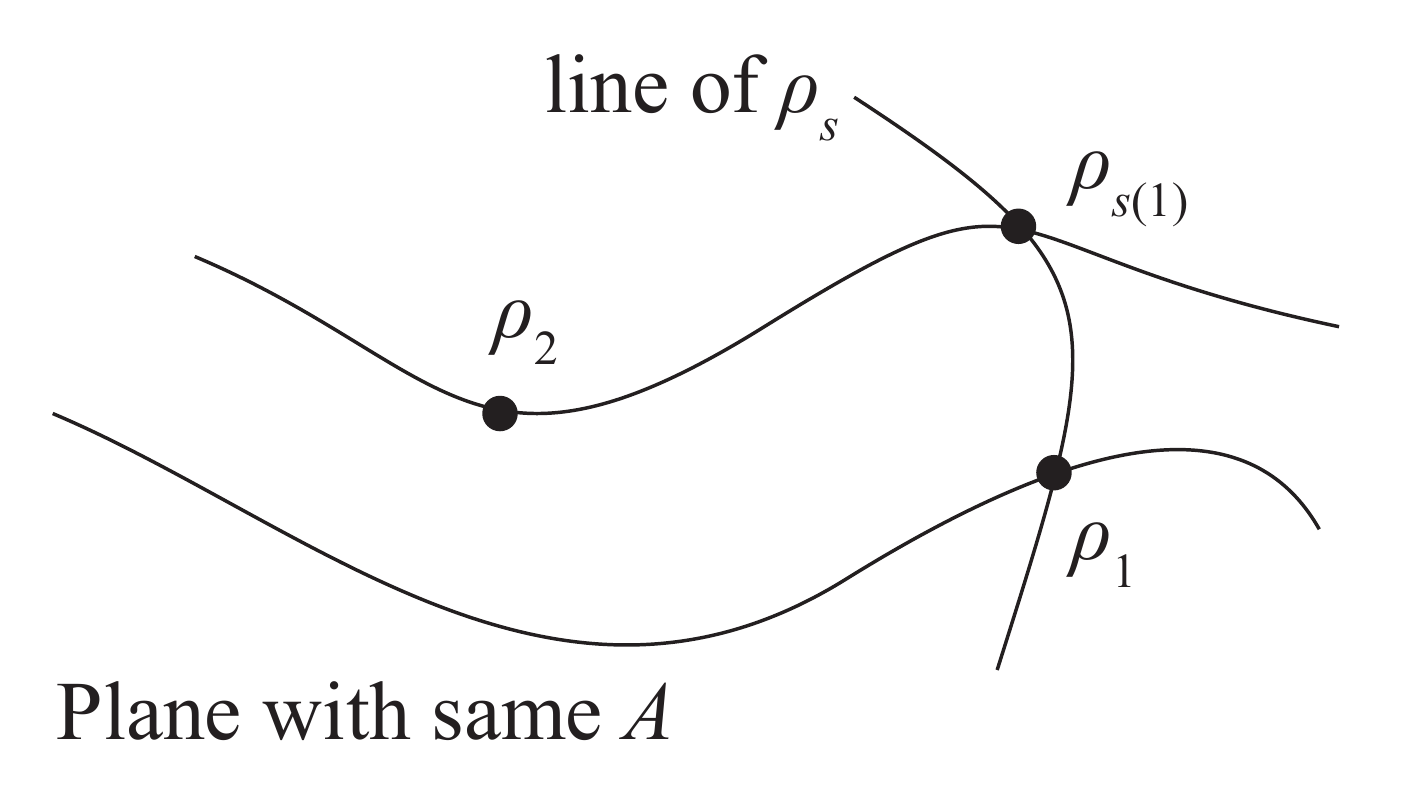}{
Schematic of the quantum information geometry.
A single point corresponds to a single density matrix.
The horizontal planes represent sets of states with the same expectation value of $A$.
The vertical line represents the trajectory of $\rho_s$ with different $s$.
The Pythagorean theorem claims that the relative entropy from $\rho_1$ to $\rho_2$ is equal to the sum of the relative entropy from $\rho_1$ to $\rho_{s(1)}$ and that from $\rho_{s(1)}$ to $\rho_2$.
}{geometry}

We finally connect the difference of energy and relative entropy.
To do this, we first discuss general properties of relative entropy.

Consider two density matrices $\rho_1$ and $\rho_2$ and an observable $A$.
Using the Pythagorean theorem in quantum information geometry~\cite{AN},  the relative entropy $D(\rho_2||\rho_1)$ is evaluated as
\balign{
D(\rho_2||\rho_1)&=D(\rho_2||\rho_{s(1)})+D(\rho_{s(1)}||\rho_1) \nt \\
&\geq D(\rho_{s(1)}||\rho_1) . \lb{Pythagorean}
}
Here, $\rho_s$ is defined as
\balign{
\rho_{s}&:=\frac{e^{\ln \rho_1 +sA}}{\Tr[e^{\ln \rho_1+sA}]}, \lb{def-rhos}
}
and $s(t)$ is a real-valued function satisfying 
\eqa{
\Tr[A\rho_{s(t)}]=(1-t)\Tr[A\rho _1]+t\Tr[A\rho _2].
}{deft}
The density matrix $\rho_s$ can be regarded as a {\it propagated} state from $\rho_1$ in the direction of $A$ (see \fref{geometry}).
We propagate the state until the expectation value of $A$ is equal to that in $\rho_2$.
The role of $t$ is re-labeling of $s$ from $0\leq s\leq s(1)$ to $0\leq t\leq 1$.
The relative entropy $D(\rho_{s(1)}||\rho_1)$ is calculated as~\cite{secondused} (see Appendix \ref{s:fisher} for its derivation)
\eqa{
D(\rho_{s(1)}||\rho_1)
=\Tr[A(\rho_2-\rho_1)]^2\int_0^1dt\frac{1-t}{J(\rho_{s(t)})}, 
}{Fisherkey}
where $J(\rho_{s})$ is the Fisher information for the family $\{\rho_{s}\}$ under the Bogoliubov-Kubo-Mori inner product \cite{AN} given by
\eq{
J(\rho_s):=\frac{\del \Tr[A\rho_s]}{\del s}.
}
The Fisher information is a kind of capacity for $A$ with respect to $s$.
If we set $A$ as the Hamiltonian of the system, $s$ as inverse temperature, and $\rho_1$ as a canonical distribution with inverse temperature $\beta$, then the Fisher information becomes the heat capacity and $\rho_s$ becomes another canonical distribution with inverse temperature $\beta-s$.

\figin{8cm}{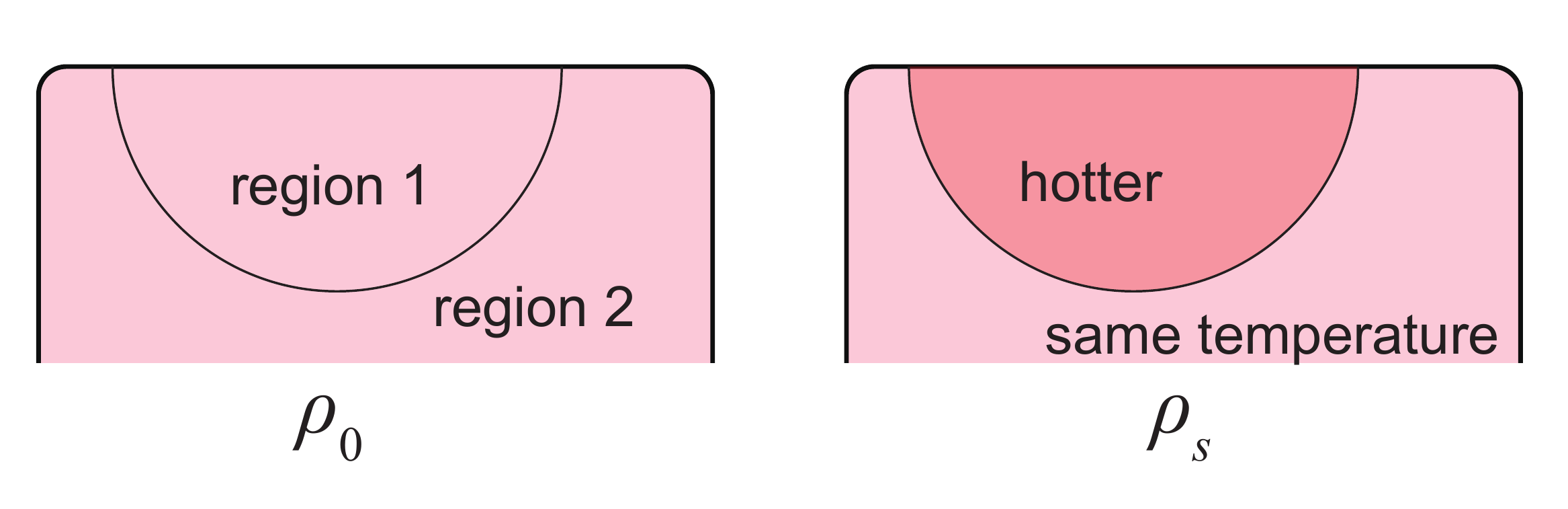}{
Schematic picture of the density matrix $\rho_s$ and the Fisher information.
The average energy in region 1 is raised, while region 2 is kept as before.
}{Fisher}

We set  $A$ in \eref{def-rhos} to $H^1$, and $\rho_1$ and $\rho_2$ to $\rho_{\rm i}^{\rm L}$ and $\rho_{\rm f}^{\rm L}$.
Although the Fisher information is not exactly equal to the heat capacity of the region 1, since the initial state of the region 1, $\rho_{\rm i}^{\rm L1}$, differs from the corresponding canonical distribution only on its boundary, the Fisher information $J(\rho_s)$ can be regarded as a {\it local} heat capacity of the region 1 (see \fref{Fisher}).
Thus, the asymptotic behavior of $J(\rho_s)$ is proportional to its volume $R^D$ in large $R$, which leads to define the renormalized Fisher information as 
\eqa{
j_{\rm max}:=\max_{R,s}\frac{J(\rho_s)}{R^D}.
}{jmax}
Insulting \eref{Fisherkey} into \eref{Pythagorean} and using monotonicity of relative entropy, we obtain
\eq{
D(\rho_{\rm f}||\rho_{\rm i})\geq D(\rho_{\rm f}^{\rm L}||\rho_{\rm i}^{\rm L})\geq \frac{(\QL )^2}{4}\int_0^1dt\frac{1-t}{J(\rho_{s(t)})}.
}
Rewriting the right-hand side of the above equation with $j_{\rm max}$ and using \eref{Rineq}, we arrive at the desired inequality~\eqref{mainnm}.

\section{Example: XX model}\lb{s:ex}

We here demonstrate how the obtained bound \eqref{mainnm} works by taking a simple example, a one-dimensional XX spin chain.
Suppose that the bath L consists of a one-dimensional discrete lattice with sites $\{ 1,2,\cdots, L\}$, where spins $1/2$ are on the sites (see \fref{XX}).
The Hamiltonian of the bath L, $H^{\rm L}$, is given by
\eq{
H^{\rm L}=-\sum_{j=1}^L J (\sigma_j^x \sigma^x_{j+1}+\sigma^y_j \sigma^y_{j+1}),
}
where $J>0$ is an interaction constant, $\sigma$ is the Pauli operator, and we identify the site $L+1$ to the site 1.
We consider that the engine is on a single site and it interacts only with the site 1.
We in particular treat a slow process (i.e., large $R$).

We first see quantities in the Lieb-Robinson bound.
In this setup, $c$ should satisfy $c\geq 1$, and $\lmd$ and $\mu$ should satisfy two constraints:
\balign{
J&\leq \lmd , \\
\frac{J}{2}&\leq \lmd e^{-\mu},
}
which are in the case of $d(x,y)=0$ and $d(x,y)=1$, respectively.
Thus, we set $c=1$, $\lmd=J$, and $\mu =\ln 2$, which implies $v=4J/\hbar$ and $\vlr=4J/\hbar\ln 2$.
In addition, $K=1$, $D=1$, $l=1$, and $H_{\rm site}^{\rm max}=J$ are satisfied in this setup.
We then find
\eq{
R^*=\frac{8J}{\hbar \ln 2}\tau +C
}
with
\eq{
C=\frac{2}{\ln 2}\ln\( \frac{3}{\ln 2}\frac{H_{\rm op}^{\rm max}}{\QH}\) +3.
}
Let us make the operation slower and slower.
In this situation, $H_{\rm op}^{\rm max}$ decreases with the increase of $\tau$ while $\QH$ is fixed, which implies that $C$ will become negative.
If the above condition is satisfied, we have $R^*\geq 8J\tau/\ln 2$.

\figin{5cm}{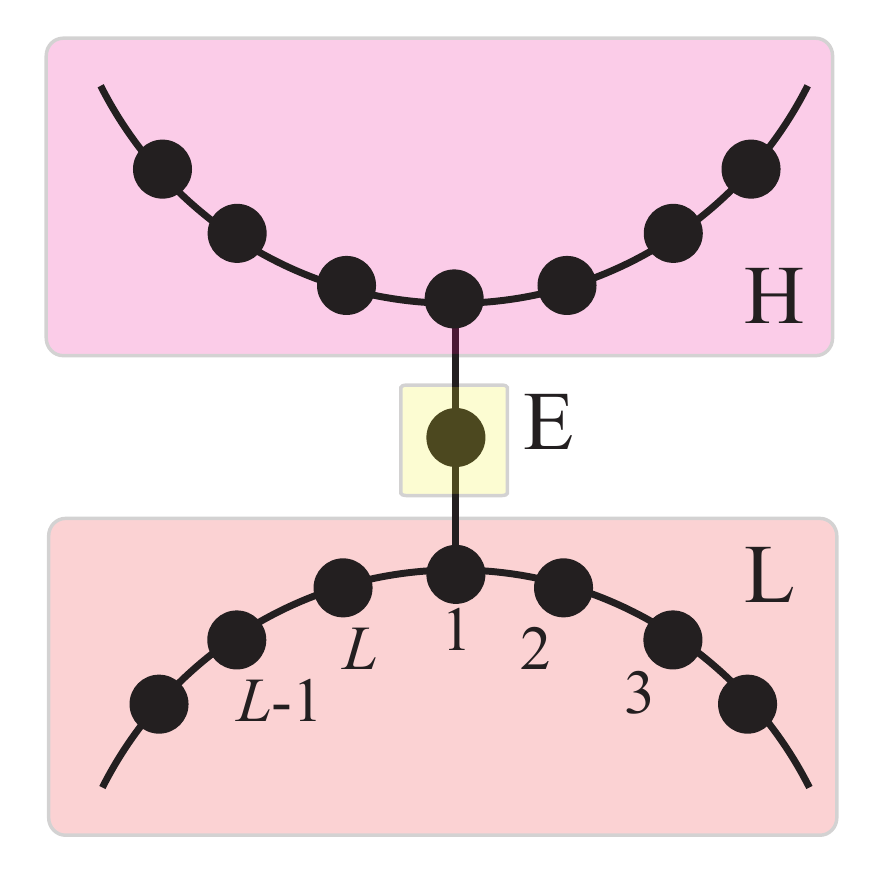}{
Schematic of the analyzed XX model.
The bath L is a periodic one-dimensional chain with length $L$.
The engine is a single site and it interacts with the bath L only through the site 1. 
}{XX}

We next see the Fisher information.
The initial state of the bath L is written as $\rhol_{\rm i} =e^{-\beta H^{\rm L}}/\Tr [e^{-\beta H^{\rm L}}]$.
This density matrix is solvable through the Jordan-Wigner transformation, which maps this system to a free fermion system.
By approximating the Fisher information by the variance of $H^1$ with $\rho_0$ and neglecting oscillating terms with high frequency, $j_{\rm max}$ is estimated above as (Derivation is shown in Appendix.\ref{der-j-XX})
\eqa{
j_{\rm max}\leq  \frac{8\bL J^3}{(1+e^{-\bL J})^2}.
}{j-XX}

Combining them, the simplified version of our main inequality \eqref{mainnm} reads 
\eq{
\eta \leq \etac -\frac{\hbar \ln 2(\QL)^2(1+e^{-\bL J})^2}{256 (\bL)^2 \QH J^3}\frac{1}{\tau}.
}
We can easily see the finiteness of the coefficient of $1/\tau$.

\section{Transient case}\lb{s:transient}

In this section, we consider the case of transient processes with multiple baths.
We again assume that there is no interaction between the system and the baths at the initial and the final state, and the initial states of the heat baths are in canonical distributions.
We shall evaluate the entropy production of the total systems given by
\eq{
\langle \hat{\sigma} \rangle :=\sum_j \beta _j\Tr [\hat{H}_j (\rho ^j_{\rm f}-\rho ^j_{\rm i})]+\Di S(\rho ^{\rm E}).
}
Here, $\rho ^j$ represents the reduced density matrix of $\rho$ to the $j$-th bath, and $\Di S(\rho ^{\rm E}):=S(\rho ^{\rm E}_{\rm f})-S(\rho ^{\rm E}_{\rm i})$ is the difference of the von Neumann entropy of the engine.

In a manner similar to Sec.~\ref{s:outline}, we obtain
\balign{
&D(\rho _{\rm f}||\rho _{\rm i}) \nt \\
=&-S(\rho _{\rm i})-\Tr [\rho _{\rm f}\ln \rho _{\rm i}]  \nt \\
=&\Tr [\rho _{\rm i}^E\ln \rho _{\rm i}^E]+\sum_j\Tr [\rho _{\rm i}^j\cdot (-\beta^j H^j)]  -\Tr [\rho _{\rm f}^E\ln \rho _{\rm i}^E] \nt \\
&-\sum_j\Tr [\rho _{\rm f}^j\cdot (-\beta^j H^j)] \nt \\
=&\sum_j \beta _j\Tr [ H^j (\rho ^j_{\rm f}-\rho ^j_{\rm i})]+S(\rho ^{\rm E}_{\rm f})-S(\rho ^{\rm E}_{\rm i})+D(\rho _{\rm f}^E||\rho _{\rm i}^E), \lb{trans-mid}
}
where $H^j$ is the Hamiltonian of the $j$-th bath.
The subadditivity of the von Neumann entropy~\cite{NC} suggests
\balign{
D(\rho _{\rm f}||\rho _{\rm i})-D(\rho _{\rm f}^E||\rho _{\rm i}^E)
=&S(\rho _{\rm f}^E)-S(\rho _{\rm f})-\sum_j \Tr [\rho _{\rm f}^j\ln \rho _{\rm i}^j] \nt \\
\geq &-\sum_j S(\rho _{\rm f}^j)-\sum_j \Tr [\rho _{\rm f}^j\ln \rho _{\rm i}^j] \nt \\
=&\sum_jD(\rho _{\rm f}^j||\rho _{\rm i}^j),
}
which directly implies
\eq{
\la \hat{\sigma} \ra \geq \sum_jD(\rho _{\rm f}^j||\rho _{\rm i}^j)
}
with the aid of \eref{trans-mid}.
The right-hand side can be evaluated in a manner similar to that in the previous section.

Let us consider an isothermal process in $0\leq t\leq \tau$ with heat emission $Q$ to the bath as an example.
In this case, the conventional second law is equivalent to the principle of maximum work:
\eq{
W\leq -\Di F
}
where the extracted work $W$ is given in \eref{def-work} and the difference of the Helmholtz free energy $\Di F$ is defined as
\eq{
\Di F:=\Tr[H^{\rm E}(\tau)\rho_{\rm f}^{\rm E}]-\Tr[H^{\rm E}(0)\rho_{\rm i}^{\rm E}]+\frac{1}{\beta}(S(\rho_{\rm i}^{\rm E})-S(\rho_{\rm f}^{\rm E})).
}
Using a relation $W+\Di F=\la \hat{\sigma}\ra$, we arrive at the extended version of the principle of maximum work with finite speed operation:
\eq{
W\leq -\Di F-\frac{Q^2}{8j_{\rm max}(2\vlr \tau +C)^D} .
}

\section{Discussion}\lb{s:dis}

In this paper, we have shown a novel no-go theorem that non-Markovian quantum heat engines never attain the Carnot efficiency at finite power.
Our result relies on the non-quick energy relaxation of baths, which is rigorously characterized by the Lieb-Robinson bound.
The most basic result in this paper is \eref{info-prop}, which is applicable to any observable $A_X$, not only to conserved quantities.
This relation simply exhibits that two different operations far from a region $X$ are hard to distinguish through observation on $X$ within a short time interval.
When $A_X$ is set to a conserved quantity, this relation serves as a bound on relaxation speed.
This technique will be helpful to clarify relaxation and thermalization phenomena in a rigorous way.

It is worth noting that there exists a completely different stream to investigate a relation on efficiency and power: the problem on efficiency at maximum power.
This problem treats the efficiency of a heat engine when its power is maximum.
A phenomenological treatment within the linear response regime (i.e., endoreversible thermodynamics~\cite{endo}) shows that the efficiency at maximum power is given by the celebrated Chambadal-Novikov-Curzon-Ahlborn efficiency~\cite{Cham, Nov, CA}.
Some other universal properties are discovered in the first and second order expansion~\cite{Broeck, Esposito, ELB}, which are again obtained in the phenomenological irreversible thermodynamics.
These stream and the investigation of a trade-off relation between efficiency and speed of operation are irrelevant to each other:
Since a trade-off relation is an upper bound, it does not characterize efficiency at maximum power.
In contrast, efficiency at maximum power consider only an engine at maximum power, and thus it does not give a trade-off relation between efficiency and speed which covers, of course, engines not at the maximum power.
In particular, the information on efficiency at maximum power does not exclude the compatibility of the Carnot efficiency and finite power.

We remark that we consider only the contribution from the bath L to the entropy production for a simple expression of the efficiency.
We can evaluate that from the bath H in a similar manner, and by taking its contribution into account we obtain the following stronger bound:
\balign{
\eta\leq \eta_{C}&-\frac{(Q^{\rm L})^2}{8\beta^{L}Q^{\rm H} j_{\rm max}(2\vlr \tau+C)^D}  \nt \\
&-\frac{Q^{\rm H}}{8\beta^{L} j'_{\rm max}(2\vlr ' \tau+C')^D} ,
}
where $ j'_{\rm max}$, $\vlr '$ and $C'$ are defined for the bath H in a similar manner to $j_{\rm max}$, $\vlr$, and $C$.

We also remark on our setup of lattice systems.
We restrict our attention to lattice systems, which allows rigorous evaluation of the speed of energy spread by the Lieb-Robinson bound.
Although some baths are not described as lattice systems, we consider that the restriction to lattice systems is only for rigorous evaluation and our essential idea still holds for non-lattice baths.
In fact, if one succeed in extending the Lieb-Robinson bound to non-lattice systems, our result is directly extended to such non-lattice systems.

Our result still provides meaningful information in thermodynamic limit.
We suppose that the time interval of an operation $\tau$ is scaled by the size of the engine $O(V^{1/D})$, where $V$ is the volume of the engine.
Since both $\QH$ and $\QL$ are of order $O(V)$ and $C$ is of order $O(l)=O(V^{1/D})$, we find that the second term of the left-hand side of \eref{mainnm} is $O(1)$ and \eref{mainnm} is still a meaningful inequality in the thermodynamic limit.
In this case, the obtained bound corresponds to the bound of efficiency in terms of exergy, which is a bound for the case with a finite size bath and is less than the Carnot efficiency.

\acknowledgments

We appreciate Keiji Saito, who we think almost as a co-author, for fruitful discussion and many helpful comments.
We also appreciate Eiki Iyoda and Kaoru Yamamoto for careful reading and useful comments.
NS is grateful to Kazuya Kaneko for his readable seminar on the Lieb-Robinson bound.
NS was supported by Grant-in-Aid for JSPS Fellows JP17J00393.

\appendix

\section{Expression with work storage}\lb{storage}

In the present paper, we treat the work storage implicitely.
However, recently, the explicit treatment of the work storage is studied actively \cite{TH15, Brandao, HO, SSP, TWO, oneshot1, oneshot2, oneshot3, Car2, Popescu2015, review, catalyst, Malabarba, m-based, T-W, MTH, Tasaki2015}. 
It is noteworthy that our results are valid perfectly under the explicit treatments of the work storage.
In this Appendix, we introduce the setup for the explicit treatment, and show that our results are indeed applicable to it.

\figin{5cm}{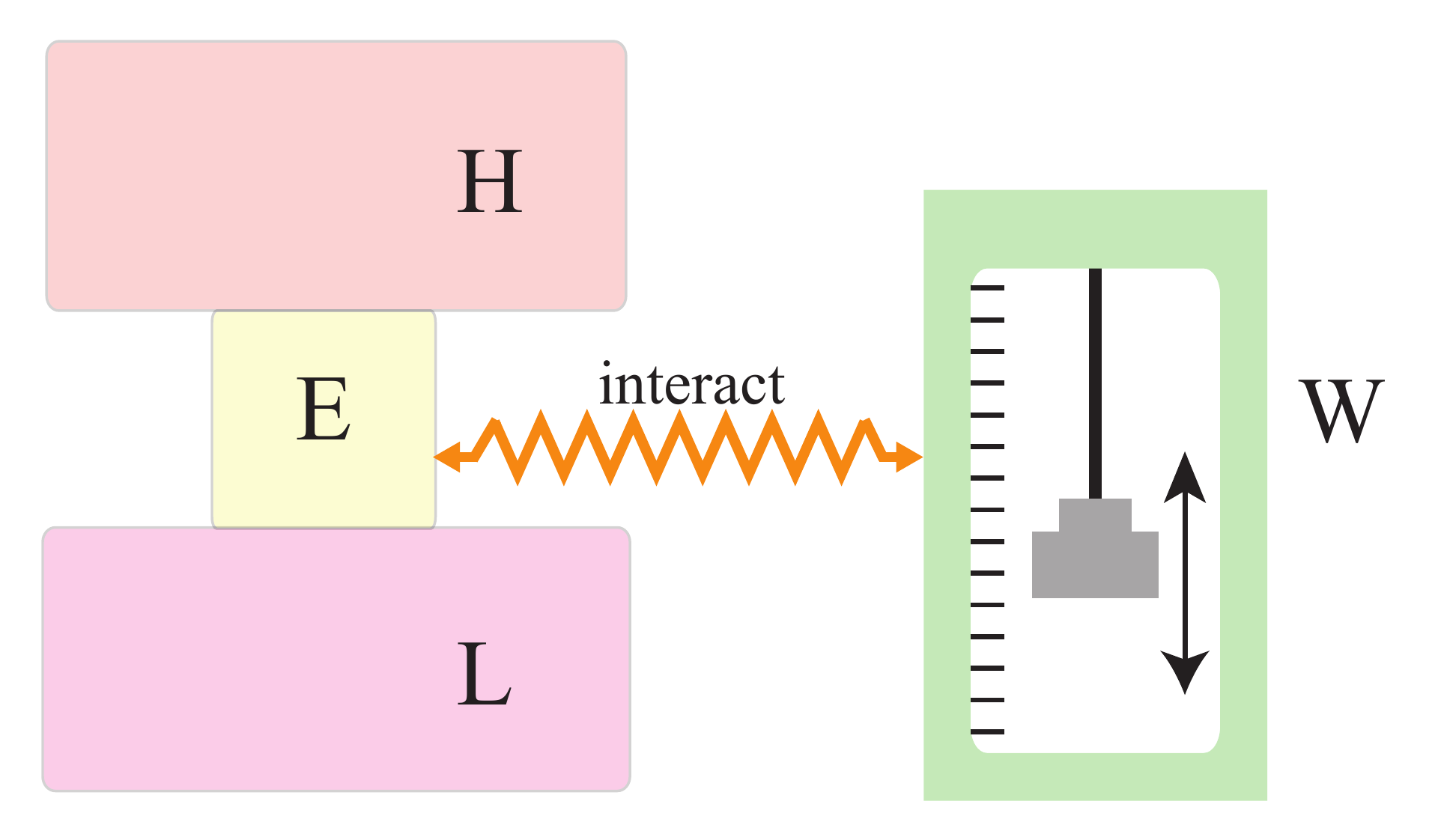}{
Schematic of a system with work storage W, which interacts only with E, not H or L.
The work storage stores extracted work as in the usable form of internal energy.
}{storage}

We firstly rewrite our setup.
The engine E and the two heat baths H and L are the same as the main text.
Namely, we prepare a quantum lattice system, and refer to the subsets of the sites $\lmde$, $\lmdh$, $\lmdl$ as the engine, the hot bath, and the cold bath, respectively.
$\lmde$ is attached to both $\lmdh$ and $\lmdl$, and $\lmdh$ is not attached to $\lmdl$.
We also define $\tle$ as a set of sites in $\lmde$ and its nearest-neighbor ones.
We then add the work storage W to the setup (see \fref{storage}).
The work storage W is not necessarily a lattice system.
$\lmde$ is also attached to W, and  neither $\lmdh$ nor $\lmdl$ is attached to W.

Next, we explain the dynamics.
We refer to the Hamiltonian of the total system including the work storage as $\htot'(t)$.
We assume that $\htot'(t)$ is written as follows:
\begin{align}
\htot'(t)=\htot(t)+H^{\rm EW}(t)+\hw .
\end{align}
Here, $\htot$ is the Hamiltonian of E, H, and L.
As in the main text, we assume that $\htot$ is the sum of one-body Hamiltonians and nearest-neighbor two-body interaction Hamiltonians.
$\hw$ is the Hamiltonian of W, and $H^{\rm EW}(t)$ is the interaction between $W$ and $E$.
The support of $H^{\rm EW}(t)$ is only on $\lmde$ and W. 
In a cyclic process in $0\leq t\leq \tau$, the Hamiltonian is changed with time as satisfying $\htot (0)=\htot (\tau)$ and $H^{\rm EW} (0)=H^{\rm EW} (\tau)$.
Same as the main part, the Hamiltonian $\htot(t)$ is time-dependent only on $\tle$.
Hence, the Hamiltonian $\htot'(t)$ is time-dependent only on $\tle$ and $W$, and can be decomposed into seven parts as
\balign{
\htot (t)=&H^{\rm E}(t)+H^{\rm EH}(t)+H^{\rm EL}(t)+H^{\rm EW}(t) \nt  \\
&+H^{\rm H}+H^{\rm L}+\hw,
}
where $H^X$ ($X=$ E, H, L) acts only on $X$, and $H^{{\rm E}X}$ ($X=$ H, L) is the sum of all interaction Hamiltonians between a site in $\lmde$ and that in $\Lmd_X$.
The Hamiltonians of the baths and the work storage are set to be time-independent because we consider the situation that the external operation acts only on the engine and the bath is not driven as explained above.
We also suppose that the engine is initially separated from baths and work storage: $H^{\rm EH}(0)=H^{\rm EL}(0)=H^{\rm EW}(0)=0$.
This setup corresponds to the situation that the initial state of the engine and that of the baths (i.e., canonical distributions) are prepared independently.
The Hamiltonian $\htot'(t)$ causes the following unitary time evolution:
\begin{align}
U_{\mathrm{tot'}}:=T\exp[-\int^{\tau}_{0}\htot'(t)dt],
\end{align}
where $T$ represents the time-ordered product.
We require that $U_{\mathrm{tot'}}$ satisfies the law of energy conservation:
\begin{align}
[U_{\mathrm{tot'}},\htot'(0)]=0.
\end{align}
If you want, we can also require some extra conditions which make the extracted energy in work storage ``work.''
This extra conditions change depending on the ``definition of work'' that we choose. 
However, as we will show in the end of this section, our results will be valid whichever definition we choose.

Let us show the definition of work and heat.
We denote the density matrix of the total system at time $t$ by $\rtots (t)$, and write $\rho_{\rm i}:=\rtots (0)$ and $\rho_{\rm f}:=\rtots (\tau)$.
The partial trace of $\rho$ to $X$ ($X=$ E, H, L, W) is denoted by $\rho^X$.
Using the canonical distribution of baths $X$ expressed as $\rhoc{X}:=e^{-\beta^XH^X}/Z^X$ ($X=$H,L) with $Z^X:=\Tr[e^{-\beta^XH^X}]$, we set the initial state as a product state $\rho_{\rm i}=\rhoE_{\rm i}\otimes \rhoc{\rm H}\otimes\rhoc{\rm L}\otimes \rho^{\rm W}_{\rm i}$, where $\rhoE_{\rm i}$ is arbitrary.
$\rho^{\rm W}_{\rm i}$ is the initial state of W, which is discussed below.
The cyclicity of the process requires that the final state and the initial state of the engine have the same energy expectation value: $\Tr[ H^{\rm E}(0)\rhoE_{\rm f}]=\Tr[ H^{\rm E}(0)\rhoE_{\rm i}]$.
The extracted work $W$, the heat released from the bath H and that absorbed by the bath L are respectively written as
\balign{
W&:=\Tr[\hw(\tau)\rhoc{W}(\tau)]-\Tr[\hw\rhoc{W} (0)],\\
\QH&:=\Tr[H^{\rm H}(\rho^{\rm H}_{\rm i}-\rho^{\rm H}_{\rm f})], \\ 
\QL&:=\Tr[H^{\rm L}(\rho^{\rm L}_{\rm f}-\rho^{\rm L}_{\rm i})].
}
Here, the amount of the extracted work is given as the difference of the expectation value of energy in the work storage.
By adding extra conditions, our definition is equal to each of the work definitions used in Refs. \cite{TH15, HO, SSP, TWO, oneshot1, oneshot2, oneshot3, Brandao, Car2, Popescu2015, review, catalyst, Malabarba, m-based, T-W, MTH, Tasaki2015}.
For example, if we add the condition that $\rhoc{W}(\tau)$ and $\rhoc{W}(0)$ are energy pure states, our definition of work is equal to the single-shot work extraction which is treated in \cite{HO, oneshot1, oneshot2, oneshot3, Brandao, review}.
If we add the condition that the von Neumann entropy of $\rhoc{W}(\tau)$ and $\rhoc{W}(0)$ are the same, our condition is equal to the average work extraction which is used in \cite{TH15, SSP, TWO, Car2, Popescu2015,  catalyst, Malabarba, m-based, T-W, MTH, Tasaki2015}.
Regardless of which of these conditions we add, our inequality is still valid, because our inequality is a necessary condition for the possibility of the transformation without these extra conditions.
(Note that a necessary condition for the possibility of transformation with looser conditions is also necessary for the possibility of the transformation with tighter conditions.)
Therefore, our results are valid for any definition of work.

\section{Derivation of Lemma}\lb{Lemma1}

We here derive the Lemma shown in Sec.~\ref{s:info-prop}.
To compare the two operations $H_0$ and $H_0+H_{\rm op}(t)$, we divide the time interval $\tau$ into $N$ ingredients $\Di t=\tau/N$ and label $t_n:=n\Di t$.
Correspondingly, we introduce $N+1$ different Hamiltonians $H^n(t)$ ($n=0,1,\cdots ,N$) as
\eq{
H^n(t):=
\bcases{
H_0+H_{\rm op}(t) &:0\leq t\leq t_n \\
H_0 &:t_n<t\leq \tau .
}
}
We also define $\rho ^n(t)$ as the density matrix of the composite system at time $t$ evolving under the Hamiltonian $H^n(t)$ (see \fref{ope-compare}).
We finally take $N\to \infty$ limit with fixed $\tau$.

We first evaluate the change in $A_X$ between $\rho^n(\tau)$ and $\rho^{n-1}(\tau)$, which corresponds to the contribution to the change in $A_X$ from $H_{\rm op}(t)$ in $t_{n-1}\leq t<t_n$.
Defining the time evolution operator with the Hamiltonian $H_0$ for time interval $\tau-t_n$ as
\eq{
U_n:= \exp \( -\frac{i}{\hbar} (\tau -t_n)\cdot H_0\) ,
}
we obtain
\balign{
&\Tr [A_X\rho ^n(\tau )] \nt \\
=&\Tr [A_XU_{n}e^{-i/\hbar \cdot (H_0+H_{\rm op}(t_{n-1}))\Di t}\rho ^N(t_{n-1}) \nt \\
&\ \ \  \cdot e^{i/\hbar \cdot (H_0+H_{\rm op}(t_{n-1}))\Di t}U_{n}^\dagger]+O(\Di t^2) \nt \\
=&\Tr [A_X\rho ^{n-1}(\tau )]+\Tr [A_XU_{n}\frac{i}{\hbar}[\rho ^N(t_{n-1}), H_{\rm op}(t_{n-1})]U_{n}^\dagger] \Di t \nt \\
&+O(\Di t^2). \lb{speed-mid}
}
In the fourth line, we used $\rho ^N(t_{n-1})=\rho ^{n-1}(t_{n-1})$ and
\eq{
\rho ^{n-1}(t_{n-1}) +\frac{i}{\hbar}[\rho ^{n-1}(t_{n-1}), H_0] \Di t=\rho ^{n-1}(t_n)+O(\Di t^2).
}
The equation \eqref{speed-mid} suggests
\balign{
&|\Tr [A_X\rho ^{n-1}(\tau )]-\Tr[A_X\rho ^{n}(\tau )]| \nt \\
=&\frac{1}{\hbar}|\Tr [[H_{\rm op}(t_{n-1}), U_{n}^\dagger A_XU_{n}]\rho ^N(t_{n-1})]| \Di t+O(\Di t^2) \nt \\
\leq&\frac{1}{\hbar}\|[H_{\rm op}(t_{n-1}), U_{n}^\dagger A_XU_{n}]\| \Di t+O(\Di t^2) , \lb{speed-mid1.5}
}
where we used the cyclic property of the trace in the second line and a relation $\| \rho \| =1$ in the third line.
By noting that $H_{\rm op}(t_{n-1})$ is an operator on $\tle$, and $U_{n}^\dagger A_XU_{n}$ is an operator on $X$ which evolves during time interval $\tau-t_n$, the Lieb-Robinson bound \eqref{LR} implies
\balign{
&\frac{1}{\hbar}\|[H_{\rm op}(t_{n-1}), U_{n}^\dagger A_XU_{n}]\| \Di t+O(\Di t^2) \nt \\
\leq& \frac{c}{\hbar}\| H_{\rm op}(t_{n-1})\| \| A_X\| |\tle||X|e^{-\mu d({\rm \tilde{E},X})} \( e^{v(N-n)\Di t}-1\) \Di t \nt \\
&+O(\Di t^2), \lb{speed-mid2}
}
where $c$, $\mu$, $\nu$ take the same values as in the Lieb-Robinson bound \eqref{LR}.
Combining Eqs.~\eqref{speed-mid1.5} and \eqref{speed-mid2}, and summing them from $n=1$ to $N$ and taking the limit $N\to \infty$, we finally obtain the desired relation
\balign{
&|\Tr [A_X\rho ^0(\tau )]-\Tr[A_X\rho ^N(\tau )]| \nt \\
\leq& \int _0^\tau dt\frac{c}{\hbar}\| H_{\rm op}(t)\| \| A_X\| |\tle||X|e^{-\mu d({\rm \tilde{E},B})} \( e^{v(\tau -t)}-1\) \nt \\
\leq& \frac{c}{\hbar}H_{\rm op}^{\rm max} \| A_X\| |\tle||X|e^{-\mu d({\rm \tilde{E},B})} \( \frac{e^{v\tau}-1}{v}-\tau \) ,
}
where we defined $H_{\rm op}^{\rm max} :=\max_{t}\| H_{\rm op}(t)\|$.
This is the desired bound.

\figin{7cm}{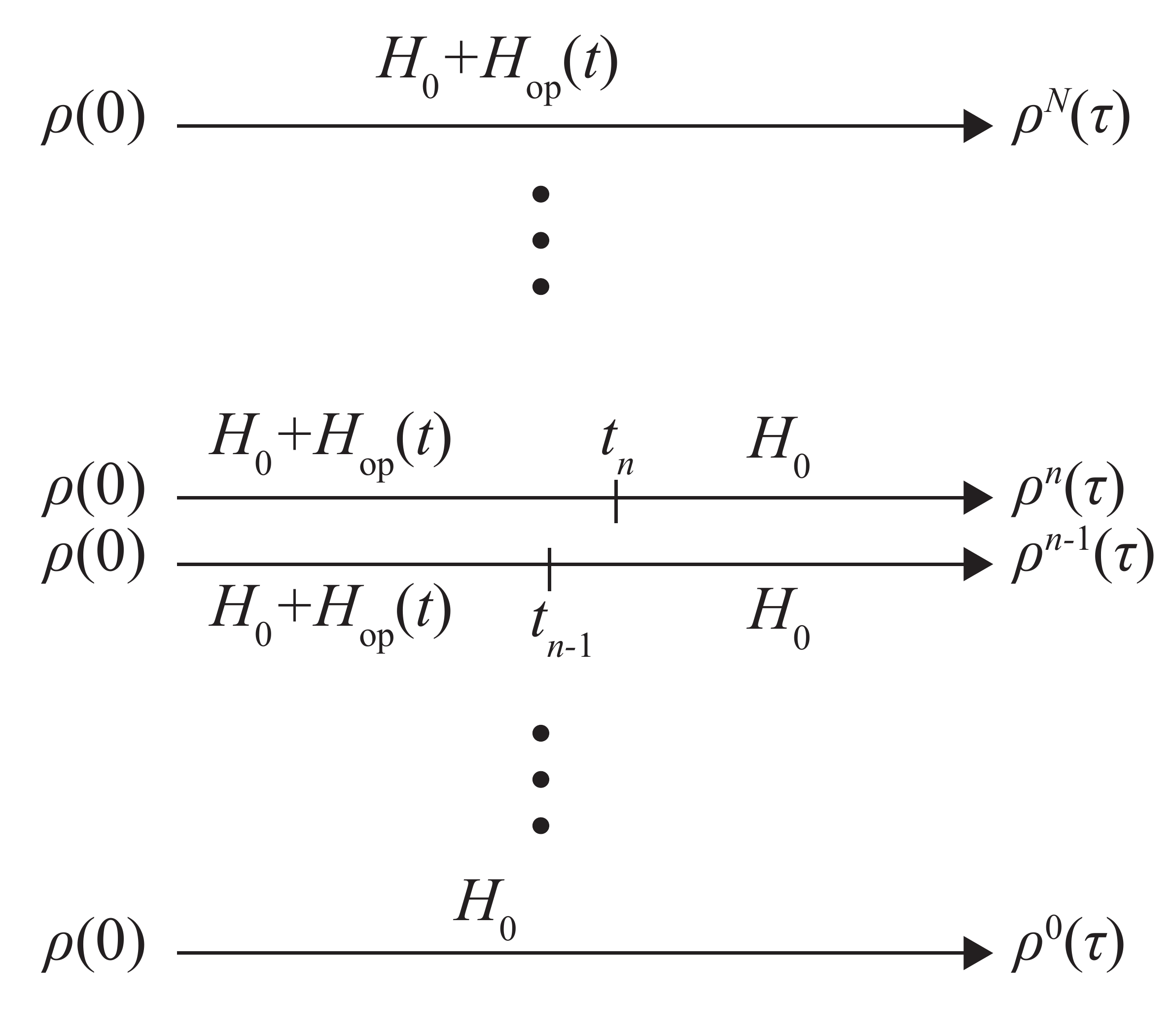}{
Schematic image of the proof of \eref{info-prop}.
We compare the expectation value of $A_X$ between $\rho^N(\tau)$ and $\rho^0(\tau)$ by introducing $N-1$ novel Hamiltonians $H^n(t)$ and density matrices $\rho^n(t)$ and piling up the difference of that between $\rho^n(\tau)$ and $\rho^{n-1}(\tau)$.
}{ope-compare}

\section{Derivation of \eref{DiE1}}\lb{derive-DiE}

We here evaluate the upper bound for energy increase outside of the region 1.
We set $R$ larger than $(D-1)/\mu+1$.
We set $A_X$ in \eref{info-prop} as a local Hamiltonian and sum up this inequality for all local Hamiltonians whose support has an overlap with the region 2.
We then obtain
\balign{
&\Di E_{12}+\Di E_2 \nt \\
\leq& \sum_{r=R}^\infty kvr^{D-1}e^{-\mu r}\( \frac{e^{v\tau}-1}{v}-\tau \)  \nt \\
\leq& ke^{v\tau}\int_{R-1}^\infty dr r^{D-1}e^{-\mu r} \nt \\
=&ke^{v\tau}\frac{e^{-\mu (R-1)}}{\mu}\sum_{i=0}^{D-1}\frac{1}{\mu^i}\frac{(D-1)!}{(D-1-i)!}(R-1)^{D-1-i} \nt \\
\leq&ke^{v\tau}\frac{e^{-\mu (R-1)}}{\mu}\sum_{i=0}^{D-1}\frac{2^i}{\mu^i}\frac{(D-1)!}{(D-1-i)!}(R-1)^{D-1-i} \nt \\
\leq& ke^{v\tau}\frac{e^{-\mu (R-1)}}{\mu}\sum_{i=-\infty}^{D-1}\frac{2^i}{\mu^i}\frac{(D-1)!}{(D-1-i)!}(R-1)^{D-1-i} \nt \\
=& ke^{v\tau}\frac{e^{-\mu (R-1)/2}2^{D-1}(D-1)!}{\mu^D} ,
}
where we defined 
\eqa{
k:=\frac{K{c}}{v\hbar} H_{\rm op}^{\rm max} H_{\rm site}^{\rm max}2|\tle| e^{\mu l}.
}{defk}
In the second line, we used the fact that the number of sites whose distance from the site $y$ is $r$ is less than $Kr^{D-1}$.
In the last line, we used the Taylor expansion of the exponential function.
The constants $K$ and $D$ are given in \eref{def-dim}.
The condition $\Di E_{12}+\Di E_2\leq \QL/2$ directly implies the desired condition for $R$.

\section{Derivation of \eref{Fisherkey}}\lb{s:fisher}

We here demonstrate the derivation of \eref{Fisherkey}.
By inserting the definition of $\rss$, the relative entropy $D(\rss ||\rho_1)$ is calculated as
\eq{
D(\rss ||\rho_1) =s(1) \Tr [\rss A]-\mu (s(1)),
}
where we defined $\mu (s):=\ln \Tr [e^{\ln \rho_1 +sA}]$.
The differentiation of \eref{deft} with respect to $t$ is written as
\eq{
\frac{ds}{dt}\frac{d\Tr [A\rho_s]}{ds}=\Tr [A\rss]-\Tr [A\rho_1],
}
which leads to the transformation of a differential form with $t$ to that with $s$ as
\eq{
dtJ^{-1}(\rst )=\frac{1}{\Tr [A\rss]-\Tr [A\rho_1]}ds.
}
We define the inverse function of $s(t)$ (given in \eref{deft}) as $t(s)$, which satisfies
\eq{
1-t(s)=-\frac{\Tr [A\rho_s]-\Tr [A\rss]}{\Tr [A\rss]-\Tr [A\rho_1]}.
}
Using both the above relation and the following relation
\eq{
\int_0^{s(1)}ds\Tr [A\rho_s]=\mu (s(1)),
}
we arrive at \eref{Fisherkey}:
\balign{
&(\Tr [A\rss]-\Tr [A\rho_1])^2\int_0^1 dt\frac{1-t}{J(\rst )} \nt \\
=&\int_0^{s(1)}ds(\Tr [A\rss]-\Tr [A\rho_s]) \nt \\
=&D(\rss ||\rho_1).
}

\section{Derivation of \eref{j-XX}}\lb{der-j-XX}

In this Appendix, we evaluate the normalized Fisher information $j_{\rm max}$ in the XX model discussed in \sref{ex}.

First, using the Jordan-Wigner transformation, this system is mapped to a free fermion system with the Hamiltonian 
\eq{
H=-\sum_{j=1}^L \frac{J}{2}(\cdg_{j+1}c_j+\cdg_jc_{j+1}),
}
where $\cdg_j$ and $c_j$ are the creation and annihilation operator of a fermion at the site $j$.
This Hamiltonian is known to be diagonalized as~\cite{Taka}
\eq{
H=-\sum_{q=0}^{L-1}J\cos \phi_q \cdot \cdg _qc_q,
}
where we defined the wave number $\phi_q :=2\pi q/L$ ($q=0,1,\cdots ,L-1$) and fermion operators
\balign{
\cdg _q&:=\frac{1}{\sqrt{L}}\sum_{j=1}^Le^{i\phi_q j}\cdg_j, \\
c_q&:=\frac{1}{\sqrt{L}}\sum_{j=1}^Le^{-i\phi_q j}c_j.
}
By setting the vacuum $\ket{0}$ as the state with no fermion (i.e., $c_j\ket{0}=0$ for all $j$), the energy eigenstates are obtained by operating some of the operators $\{ \cdg_q\}$ to the vacuum as $\cdg_{q_1} \cdg_{q_2}\cdots \cdg_{q_m}\ket{0}$, whose energy eigenvalue is $-\sum_{i=1}^m J\cos \phi_{q_i}$.

We set the region 1 as $\{ L-l+1, L-l+2, \cdots, L, 1,2,\cdots ,l+1\}$, and $A=J/2 \sum_{j=-l+1}^{l+1} (\cdg_{j+1}c_j+\cdg_jc_{j+1})=:H^1$, where we denote $\cdg_{-a}:=\cdg_{L-a}$.
We now use some approximations.
First, since $R$ is large, which implies small $s(1)$, we find that $J(\rho_s)$ ($0\leq s\leq s(1)$) is close to $J(\rho_0)$ and therefore approximate the former by the latter.
Second, although $H$ and $H^1$ are not commute, the commutator is small compared to $H$ and $H^1$.
This fact suggests that $J(\rho_0)$ is approximated as the variance of $H^1$:
\eq{
J(\rho_0)\simeq \Tr[(H^1)^2 \rho_0]-\Tr [H^1\rho_0]^2.
}
Note that $\rho_0=e^{-\beta H^{\rm L}}/\Tr [e^{-\beta H^{\rm L}}]$ is the canonical distribution with inverse temperature $\bL$.

Expressing $\cdg_i$ and $c_i$ in terms of $\cdg_q$ and $c_q$, we can calculate $\Tr [H^1\rho_0]$ as
\eq{
\Tr [H^1\rho_0]=2lJ\cdot \frac{1}{L}\sum_{q=0}^{L-1}\frac{\cos \phi_q}{1+e^{-\bL J \cos \phi_q}} 
}
and $\Tr[(H^1)^2 \rho_0]$ as
\begin{widetext}
\eq{
\Tr[(H^1)^2 \rho_0]
=\( 2lJ\cdot \frac{1}{L}\sum_{q=0}^{L-1}\frac{\cos \phi_q}{1+e^{-\bL J \cos \phi_q}}\) ^2
+\frac{J^2}{L^2}\sum_{j,j'=-l}^l\sum_{q,q'=0}^{L-1}\frac{\cos ((j-j')(\phi_q-\phi_{q'}))(1+\cos (\phi_q+\phi_{q'}))}{(1+e^{-\bL J \cos \phi_q})(1+e^{-\bL J \cos \phi_{q'}})}. \lb{VarH-mid}
}
\end{widetext}
In large $L$ limit, $\phi_q$ and $\phi_{q'}$ become continuous and run from 0 to $2\pi$.
Consider the situation that $\phi_q+\phi_{q'}$ is fixed and $\phi_q-\phi_{q'}$ runs from $-2\pi$ to $2\pi$.
If $\abs{j-j'}>\bL J$, the numerator $\cos ((j-j')(\phi_q-\phi_{q'}))$ oscillates quickly compared to its denominator $(1+e^{-\bL J \cos \phi_q})(1+e^{-\bL J \cos \phi_{q'}})$, and the summation turns to be negligible.
We thus find
\begin{widetext}
\balign{
\Tr[(H^1)^2 \rho_0]-\Tr [H^1\rho_0]^2
\simeq&\frac{J^2}{L^2}\sum_{\substack{j,j'=-l \\ \abs{j-j'}\leq \bL J}}^l\sum_{q,q'=0}^{L-1}\frac{\cos ((j-j')(\phi_q-\phi_{q'}))(1+\cos (\phi_q+\phi_{q'}))}{(1+e^{-\bL J \cos \phi_q})(1+e^{-\bL J \cos \phi_{q'}})} \nt \\
\leq&4l\bL J\frac{J^2}{L^2}\sum_{q,q'=0}^{L-1}\frac{1+\cos (\phi_q+\phi_{q'})}{(1+e^{-\bL J \cos \phi_q})(1+e^{-\bL J \cos \phi_{q'}})} \nt \\
\leq&4l\bL J \frac{2J^2}{(1+e^{-\bL J})^2},
}
\end{widetext}
which implies the desired \eref{j-XX}.

\section{Case of general Hamiltonian}\lb{s:gen}

\figin{7cm}{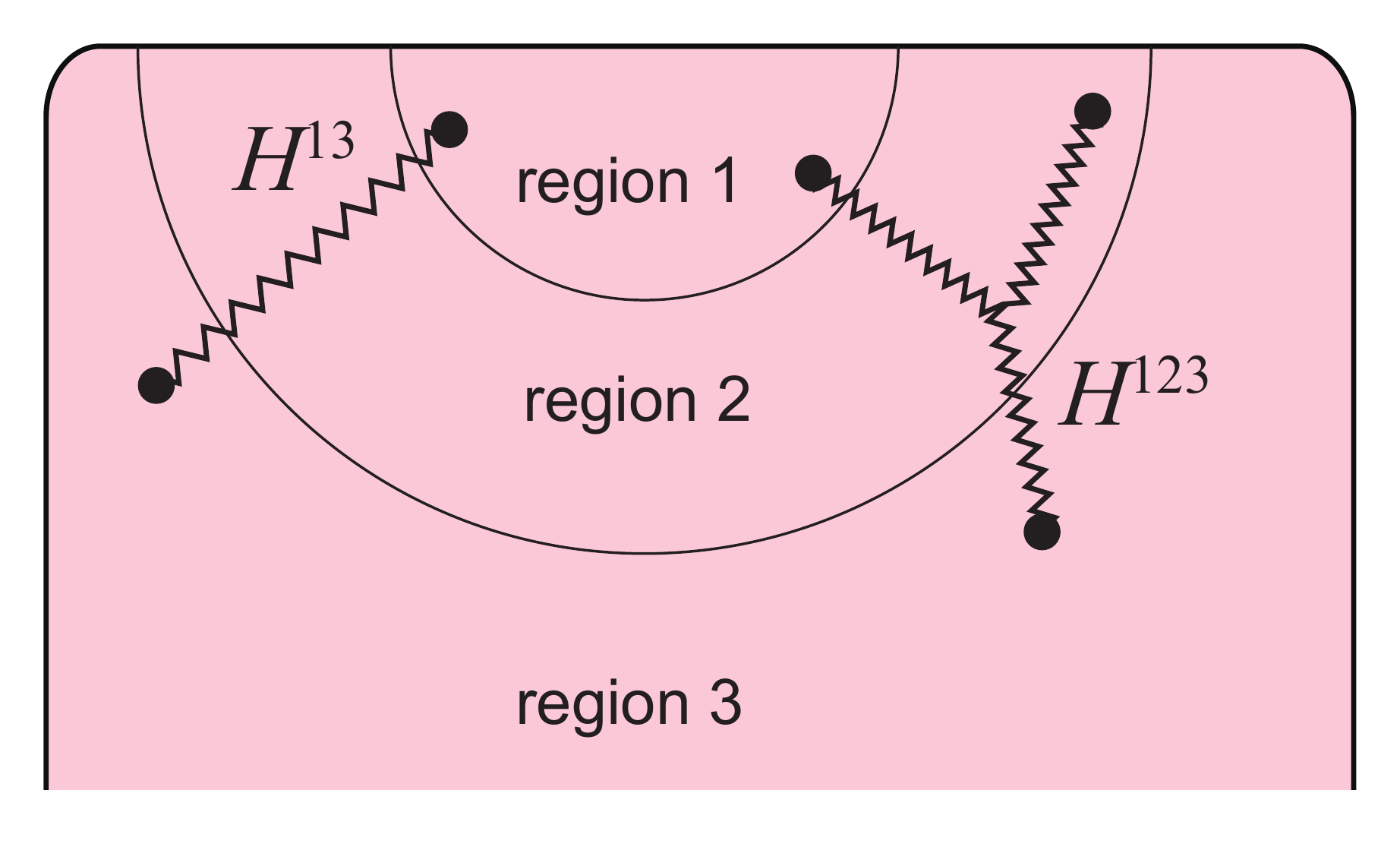}{
Schematic of the division of the bath into the region 1,2, and 3, and examples of local Hamiltonians in $H^{13}$ and $H^{123}$.
}{genH}

In this Appendix, we consider the case with general exponentially-decaying local Hamiltonians (i.e., those satisfying \eref{expdecay}) not restricted to nearest-neighbor interactions.
In this case, we divide the bath L into three regions (see \fref{genH}): 
\bi{
\item Region 1: sites $y$ with ${\rm dist}(x,y)\leq R$
\item Region 2: sites $y$ with $R<{\rm dist}(x,y)\leq 2R$
\item Region 3: sites $y$ with $2R+1\leq {\rm dist}(x,y)$.
}
Correspondingly, we decompose the Hamiltonian on L into seven parts, $H^1$, $H^2$, $H^3$, $H^{12}$, $H^{23}$, $H^{13}$, and $H^{123}$, which are the sum of local Hamiltonians with its support on only $\Lmd^1$, $\Lmd^2$, $\Lmd^3$, among $\Lmd^1$ and $\Lmd^2$, among $\Lmd^2$ and $\Lmd^3$, among $\Lmd^1$ and $\Lmd^3$, and among $\Lmd^1$ and $\Lmd^2$ and $\Lmd^3$, respectively.
We also decompose the total energy into seven parts, $E_1, E_2, E_3, E_{12}, E_{13}, E_{23}, E_{123}$, where the former three energies are in the region 1$\sim$3, and the middle three energies are many-body interactions with sites among two regions (1,2), (1,3), and (2,3), and the last energy is many-body interactions with sites among three regions (1,2,3).

We first evaluate the change in energy inside  the composite region of 2 and 3, $\Di E_{\rm in}^{23}:=|\Di E_2|+|\Di E_3|+|\Di E_{23}|$, from $t=0$ to $\tau$.
By following argument similar to that in \sref{info-prop}, the maximum of the change in energy in these regions is evaluated as
\eqa{
\Di E_{\rm in}^{23} \leq \sum_{r=R+1}^\infty k'v r^{D-1}e^{-\mu r} \( \frac{e^{v\tau}-1}{v}-\tau \) .
}{Ein23}
Here, the coefficient $k'$ is set to
\eq{
k':=\frac{Kc}{\hbar v}H_{\rm op}^{\rm max} H_{\rm site}^{\rm max} |\tle| N_{\rm int}e^{\mu l} 
}
corresponding to \eref{defk}, where we defined $N_{\rm int}$ such that the interaction in the bath is at most $N_{\rm int}$-body interaction.

We next evaluate the change in energy of many-body interactions among sites in both region 1 and 3 (i.e., $E_{13}+E_{123}$).
To obtain this, we first evaluate the sum of the operator norm of $H_{13}$ and $H_{123}$ as follows:
\balign{
\| H_{13}\| +\| H_{123}\| 
\leq &\sum_{x\in {\Lmd^1}}\sum_{y\in {\Lmd^3}}\sum_{Z\ni x,y}\| h_Z\| \nt \\
\leq &K\sum_{x=1}^R x^{D-1}\lmd\sum_{r=R+1}^\infty Kr^{D-1} e^{-\mu r} \nt \\
\leq &K \int_0^{R+1}x^{D-1}dx \cdot \lmd\sum_{r=R+1}^\infty Kr^{D-1} e^{-\mu r} \nt \\
=&K \frac{(R+1)^D}{D} \cdot \lmd\sum_{r=R+1}^\infty Kr^{D-1} e^{-\mu r}. \lb{openorm13} \nt \\
}

Setting $R\geq (D-1)/\mu $, we obtain
\begin{widetext}
\balign{
\Di E_{\rm out}^{12}
:=&|\Di E_3|+|\Di E_{13}|+|\Di E_{23}|+|\Di E_{123}| \nt \\
\leq& \Di E_{\rm in}^{23}+| \Di E_{13}|+ |\Di E_{123}| \nt \\
\leq& \sum_{r=R+1}^\infty k'e^{v\tau} r^{D-1}e^{-\mu r} +K^2 \frac{(R+1)^D}{D} \lmd\sum_{r=2R+1}^\infty r^{D-1} e^{-\mu r} \nt \\
\leq&  \frac{(R+1)^D}{D} \sum_{r=R+1}^\infty k'e^{v\tau} r^{D-1}e^{-\mu r} +K^2 \frac{(R+1)^D}{D} \lmd\sum_{r=R+1}^\infty r^{D-1} e^{-\mu r} \nt \\
\leq& \( k'e^{v\tau} +K^2 \lmd\)  \frac{(R+1)^D}{D} \int_{R}^\infty dr r^{D-1}e^{-\mu r} \nt \\
\leq&\( k'e^{v\tau} +K^2  \lmd\)  \frac{1}{D} \frac{e^{-\mu R}}{\mu}\sum_{i=0}^{D-1}\frac{1}{\mu^i}\frac{(D-1)!}{(D-1-i)!}R^{D-1-i} \nt \\
\leq&\( k'e^{v\tau} +K^2  \lmd\)  \frac{1}{D} \frac{e^{-\mu R}}{\mu}\sum_{i=-\infty}^{2D-1}\frac{2^i}{\mu^i}\frac{(2D-1)!}{(2D-1-i)!}(R+1)^{2D-1-i} \nt \\
=&(k'e^{v\tau} +K^2  \lmd)\frac{2^{2D-1}(2D-1)!}{D\mu ^{2D}}e^{-\mu (R-1)/2} .
}
\end{widetext}
In the third line, we used  $| \Di E_{13}|+ |\Di E_{123}|\leq \| H_{13}\| +\| H_{123}\|$ and \eref{openorm13}.
In the fourth line, we used ${(R+1)^D}/{D}\geq 1$.
In the fifth line, we used the fact that $r^{D-1} e^{-\mu r}$ is monotonically decreasing for $r\geq (D-1)/\mu$.
Hence, the condition $\Di E_1+\Di E_2+\Di E_{12}\geq \QL-\Di E_{\rm out}^{12}\geq \QL/2$ is satisfied for $R\geq R^*$ with
\eq{
R^*:= \frac{2}{\mu}\( \ln \( e^{v\tau}+\frac{K^2\lmd}{k'}\) -\ln \( \frac{k'D\mu ^{2D+1}\QL}{2^{2D}(2D-1)!}\) \) +1.
}
For sufficiently large $\tau$ such that $ e^{v\tau} \gg {K^2\lmd}/{k'}$, $R^*$ behaves as $2v/\mu\cdot \tau$ with a constant term.
Corresponding to \eref{mainnm}, the efficiency is bounded in terms of $R^*$ as
\eq{
\eta\leq \eta_{C}-\frac{(Q^{\rm L})^2}{8\beta^{L}Q^{\rm H} j_{\rm max}(R^*)^D}.
}

\section{Case of Markovian limit}\lb{s:markov}

\subsection{Problem and Setup}\lb{s:Markov-set}

\figin{8.5cm}{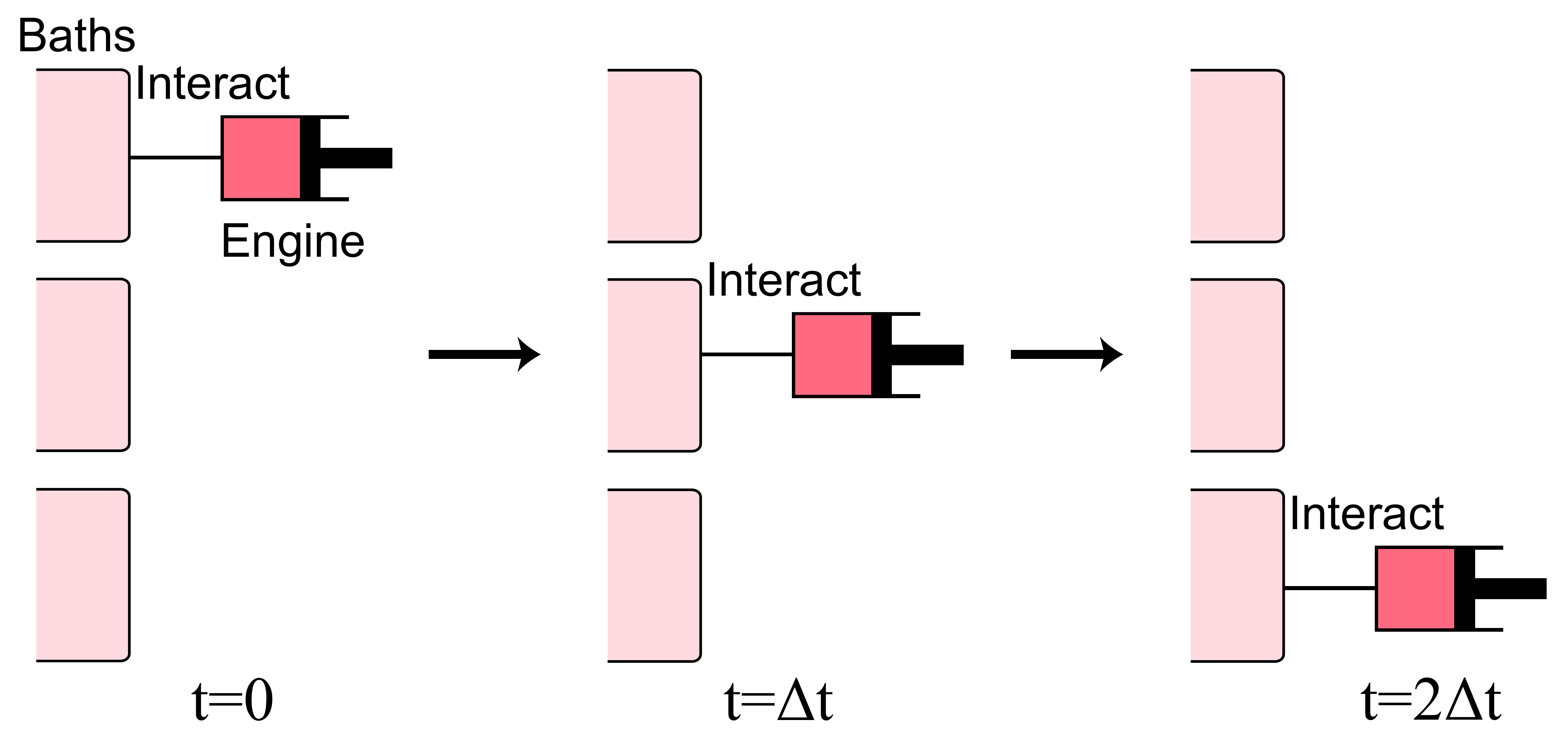}{
Schematic picture of quantum Markov processes.
We prepare infinitely many baths in canonical distribution.
We attach the engine to one of the baths in small time interval $\Di t$, and detach the bath and attach a new bath to the system.
}{f:Markov}

Our main result~\eqref{mainnm} contains the operator norm of the local Hamiltonian of the bath L in its right-hand side in the form of the Lieb-Robinson velocity $\vlr$.
Because the operator norm of the local Hamiltonian in baths diverges and so does $\vlr$ in Markovian limit, our inequality falls down into the conventional second law in this limit.
To avoid this insufficiency, in this Appendix we demonstrate a completely different approach to exclude the possibility of coexistence of finite power and the Carnot efficiency in quantum Markov processes.

We again treat an engine with two heat baths L and H with inverse temperatures $\bH$ and $\bL$.
It is straightforward to extend our analysis to the case with three or more baths and with particle baths.
Differently from the main part on non-Markovian engines, we here do not assume that the system is on a lattice and the interaction is short-range.
The initial state of the total system is again given by $\rho_{\rm tot}(0)=\rhoE_{\rm i}\otimes \rhoc{\rm H}\otimes\rhoc{\rm L}$.
The spectrum decomposition of the initial state of the engine reads $\rhoE_{\rm i}=\sum_a p_0(a)\ket{\phi _a}\bra{\phi _a}$.
By denoting the energy eigenstates of the bath $X$ ($X=$H, L) by $\ket{\ep_{j_X}^X}$, the basis $\{ \ket{a,\jH, \jL}\} _0 :=\{ \ket{\phi _a}\ket{\ep ^{\rm H}_{\jH}}\ket{\ep ^{\rm L}_{\jL}}\}$ diagonalizes $\rho_{\rm tot}(0)$.

Suppose that the total system evolves in a small time interval from $t=0$ to $t=\Di t$ (see \fref{f:Markov}).
The density matrix of the total system at $t=\Di t$ is given by $\rtot (\Di t)=U\rtot (0)U^\dagger$ with the time evolution operator $U$.
The spectrum decomposition of the reduced density matrix reads $\Tr _{\rm H,L}[\rtot(\Di t)]=\sum_a p_{\Di t}(a)\ket{\psi _a}\bra{\psi _a}$, where the label of $a$ in $\ket{\psi _a}$ is set as satisfying $\lim_{\Di t\to 0} \braket{\psi_a |\phi _a}=1$.
Suppose that we perform a projection measurement on $\rtot (\Di t)$ with a basis $\{ \ket{a,\jH, \jL}\} _{\Di t} :=\{ \ket{\psi _a}\ket{\ep ^{\rm H}_{\jH}}\ket{\ep ^{\rm L}_{\jL}}\}$, which does not affect the evolution of the system because the engine is diagonalized with this basis and the baths are not used in the subsequent evolution of the engine as explained later.

\subsection{Quantum entropy production and trade-off relation}\lb{s:Markov-claim}

We here introduce the transition probability of a forward and its time-reversal process as
\balign{
P^{\Di t}_{a, \jH, \jL \to a',\jH', \jL'}&:=|\bra{a',\jH' ,\jL'}_{\Di t}U\ket{a,\jH,\jL}_0|^2 \\
\tilde{P}^{\Di t}_{ \tilde{a'}, \tilde{\jH'}, \tilde{\jL'} \to \tilde{a}, \tilde{\jH}, \tilde{\jL}}&:=|\bra{\tilde{a}, \tilde{\jH}, \tilde{\jL}}_0 U^\dagger \ket{\tilde{a'}, \tilde{\jH'}, \tilde{\jL'} }_{\Di t}|^2,
}
where $\ket{\tilde{a}}$ represents the time-reversal state of $\ket{a}$.
Time-reversal symmetry of unitary evolution implies 
\eqa{
P^{\Di t}_{a, \jH, \jL \to a',\jH', \jL'}=\tilde{P}^{\Di t}_{ \tilde{a'}, \tilde{\jH'}, \tilde{\jL'} \to \tilde{a}, \tilde{\jH}, \tilde{\jL}}.
}{trs}
It is known that the stochastic entropy production can be expressed as~\cite{Sagawa}
\balign{
\hat{\sigma}^{\Di t} 
:=&s(p_{\Di t}(a'))-s(p_0(a))+\sum_i\beta ^XQ^X \nt \\
=&\ln \frac{p_0(a) p^{\rm H}_{\rm can}(\jH)p^{\rm L}_{\rm can}(\jL)  P^{\Di t}_{a, \jH, \jL \to a',\jH', \jL'}}{p_{\Di t}(a')p^{\rm H}_{\rm can}(\jH')p^{\rm L}_{\rm can}(\jL') \tilde{P}^{\Di t}_{ \tilde{a'}, \tilde{\jH'}, \tilde{\jL'} \to \tilde{a}, \tilde{\jH}, \tilde{\jL}}},
}
where we defined the heat release $Q^X:=\ep ^X _{j_X}-\ep^X _{j'_X}$, the canonical distribution $p^X_{\rm can}(j_X):=e^{-\beta ^X\ep^X_{j_X}}/Z_X$ with the partition function $Z_X:=\sum_{j_X}e^{-\beta ^X\ep^X_{j_X}}$, and the stochastic entropy $s(p(a)):=-\ln p(a)$, which reproduces the Shannon entropy by taking its average.
With these definitions, both the fluctuation theorem, $\langle \exp \( {-\hat{\sigma}^{\Di t}}\) \rangle =1$, and the second law of thermodynamics, $\la \hat{\sigma}^{\Di t}\ra \geq 0$, are satisfied.

We here require a physically plausible property that in sufficiently small time interval $\Di t$ energy cannot be transported directly from a bath to the other bath.
In other words, $P^{\Di t}_{a, \jH, \jL \to a',\jH', \jL'}$ turns to be zero if both $\jH \neq \jH'$ and $\jL\neq \jL'$ hold.
We then safely define the transition rate induced by each bath as
\balign{
P^{\Di t}_{a, \jH \to a',\jH'}&:=\sum_{\jL}P^{\Di t}_{a, \jH, \jL \to a',\jH', \jL} \\
P^{\Di t}_{a, \jL \to a',\jL'}&:=\sum_{\jH}P^{\Di t}_{a, \jH, \jL \to a',\jH, \jL'}. 
}
Taking $\Di t\to 0$ limit and defining $P_{a,j_X  \to a',j'_X }:=\lim_{\Di t\to 0} P^{\Di t}_{a,j_X  \to a',j'_X }/\Di t$, we have an expression of entropy production rate $\dot{\sigma}:=\lim_{\Di t\to 0}\la \hat{\sigma}^{\Di t}\ra /\Di t$ and heat flux as
\balign{
\dot{\sigma}:=&\dot{\sigma}_{\rm H}+\dot{\sigma}_{\rm L} \\
\dot{\sigma}_X:=&\sum_{a,j_X ,a',j'_X} p_0(a) p^X_{\rm can}(j_X) P_{a,j_X  \to a',j'_X }\ln \frac{p_0(a)p^X_{\rm can}(j_X)}{p_0(a')p^X_{\rm can}(j'_X)} \\
J_Q^X :=&\sum_{a,j_X ,a',j'_X} p_0(a) p^X_{\rm can}(j_X) P_{a,j_X  \to a',j'_X } (E_{a'}-E_a),
}
Here, $\dot{\sigma}_X$ ($X=$H, L) is a part of entropy production rate contributed from the bath $X$, and $E_a:=\bra{\phi_a}H\ket{\phi_a}$ is the energy expectation value of a state $\ket{\phi_a}$, which satisfies $E_{a'}-E_a=\ep ^X _{j_X}-\ep^X _{j'_X}$ in $\Di t\to 0$ limit due to the law of energy conservation.

In line with Ref.~\cite{SST}, we can derive the following inequality for a quantum Markov engine:
\eqa{
\sum_{X={\rm H,L}} |J_Q^X|\leq \sqrt{\Theta (0)\dot{\sigma}}.
}{Markovmain}
Here, $\Theta (0)$ is defined as
\eq{
\Theta (0):=\frac{9}{8}\sum_{X={\rm H,L}}\left[ \sum_{a,j_X ,a',j'_X} (\Di E_{a'})^2 A^X _{a,j_X; a',j'_X}\right]
}
with energy fluctuation and activity:
\balign{
\Di E_{a'}:=& E_{a'}-\Tr [H\rho (0)], \\
A^X _{a,j_X; a',j'_X}:=&p_0(a)p^X_{\rm can}(j_X) P_{a,j_X  \to a',j'_X } \nt \\
&+p_0(a')p^X_{\rm can}(j_X) P_{{a}',{j}'_X \to {a},{j}_X}.
}
The derivation of \eref{Markovmain} is shown in the next subsection.

Since the bath equilibrates extremely quick compared to the time scale of the engine, we argue that throughout a process in $0\leq t\leq \tau$ the heat bath is regarded as always in canonical distribution, which implies that \eref{Markovmain} holds at any time $t$ by replacing $p_0$ and $\Theta (0)$ with $p_t$ and $\Theta (t)$.
The same assumption is seen in the Born-Markov approximation, which is used in a standard derivation of the Lindblad equation~\cite{breuer}.
Accepting this plausible requirement, we obtain the trade-off relation between power and efficiency for a cyclic process during $0\leq t\leq \tau$ with two heat baths with inverse temperatures $\bH$ and $\bL$ ($\bH<\bL$).
The integration of \eref{Markovmain} from $t=0$ to $\tau$ with the Schwarz inequality leads to $(\QH+\QL)^2\leq \tau \bar{\Theta}\Di S$, where we defined time-averaging of $\Theta$ as $\bar{\Theta}:=1/\tau \int_0^\tau dt\Theta (t)$ and entropy increase as $\Di S:=\bL\QL-\bH\QH$.
Here, the change in the entropy of the engine is zero due to cyclicity.
Then, a thermodynamic relation $\eta(\etaC-\eta)=W\DS/\{\bL(\QH)^2\}$ suggests that the work $W:=\QH-\QL$ and efficiency $\eta:=W/\QH$ satisfy
\eqa{
\frac{W}{\tau}\leq \bar{\Theta}\bL \eta (\eta_{\rm C}-\eta),
}{power-Markov}
which clearly shows that a finite power heat engine never attains the Carnot efficiency.

We remark that our result on Markovian engines is applicable to broader class of engines than the result for stationary thermoelectric transport shown in Ref.~\cite{SSarxiv}.
The result in Ref.~\cite{SSarxiv} is applicable only to stationary systems described by the Lindblad equation.
In contrast, our result also covers transient systems and stationary Markovian systems not described by the Lindblad equation (e.g., systems without the rotating wave approximation or without the weak-coupling approximation).
We note that the Markovness of the system requires that the interaction between the bath and the system is weak with respect to the bath, while it is not necessarily weak with respect to the system.
Our result is applicable to the latter one.

\subsection{Derivation of \eref{Markovmain}}\lb{s:Markov}

In this Appendix, we derive \eref{Markovmain} in line with Ref.~\cite{SST}.
We first derive
\eqa{
|J_Q^X|\leq \sqrt{\Theta_X \dot{\sigma}_X},
}{Markovkey}
from which we can easily prove \eref{Markovmain}.
We here introduce quantities
\balign{
\tilde{A}^{X,\pm} _{a,j_X; a',j'_X}:=&p_0(a)p^X_{\rm can}(j_X) P_{a,j_X  \to a',j'_X } \nt \\
&\pm p_0(a')p^X_{\rm can}(j'_X) \tilde{P}_{\tilde{a}',\tilde{j}'_X \to \tilde{a},\tilde{j}_X},
}
which consist of a probability flux $a,j_X\to a',j'_X$ and its time-reversal one $\tilde{a}',\tilde{j}'_X \to \tilde{a},\tilde{j}_X$.
We note a relation corresponding to the time-reversal invariance of escape rate in classical Markov jump processes as
\eqa{
 \sum_{\substack{a,j_X \\ (a,j_X)\neq (a',j_X')}}\tilde{P}_{\tilde{a}',\tilde{j}'_X \to \tilde{a},\tilde{j}_X}= \sum_{\substack{a,j_X \\ (a,j_X)\neq (a',j_X')}}P_{{a}',{j}'_X \to {a},{j}_X}
}{escape-trs}
for any $a', j'_X$, which follows from the following time-reversal symmetry
\eq{
P^{\Di t}_{{a},{j}_X \to {a},{j}_X}=\tilde{P}^{\Di t}_{\tilde{a},\tilde{j}_X \to \tilde{a},\tilde{j}_X}.
}
Using \eref{escape-trs}, the normalization condition and the Schwarz inequality, we obtain \eref{Markovkey}:
\begin{widetext}
\balign{
|J_Q^X|^2
=&\abs{\sum_{\substack{a.j_X, a', j'_X \\ (a,j_X)\neq (a',j_X')}}E_{a'} \( p_0(a)p^X_{\rm can}(j_X) P_{a,j_X  \to a',j'_X }-p_0(a')p^X_{\rm can}(j'_X) P_{{a}',{j}'_X \to {a},{j}_X}\) }^2 \nt \\
=&\abs{\sum_{\substack{a.j_X, a', j'_X \\ (a,j_X)\neq (a',j_X')}}E_{a'}\tilde{A}^{X,-} _{a,j_X; a',j'_X}}^2 \nt \\
=&\abs{\sum_{\substack{a.j_X, a', j'_X \\ (a,j_X)\neq (a',j_X')}}\Di E_{a'}\tilde{A}^{X,-} _{a,j_X; a',j'_X}}^2 \nt \\
=&\abs{\sum_{\substack{a.j_X, a', j'_X \\ (a,j_X)\neq (a',j_X')}}\Di E_{a'}\sqrt{\tilde{A}^{X,+} _{a,j_X; a',j'_X}}\cdot \frac{\tilde{A}^{X,-} _{a,j_X; a',j'_X}}{\sqrt{\tilde{A}^{X,+} _{a,j_X; a',j'_X}}}}^2 \nt \\
\leq&\( \sum_{\substack{a.j_X, a', j'_X \\ (a,j_X)\neq (a',j_X')}}\Di E_{a'}^2 \tilde{A}^{X,+} _{a,j_X; a',j'_X} \) \( \sum_{\substack{a.j_X, a', j'_X \\ (a,j_X)\neq (a',j_X')}} \frac{\( \tilde{A}^{X,-} _{a,j_X; a',j'_X}\) ^2}{\tilde{A}^{X,+} _{a,j_X; a',j'_X}}\) \nt \\
\leq&\( \sum_{\substack{a.j_X, a', j'_X \\ (a,j_X)\neq (a',j_X')}}\Di E_{a'}^2 \tilde{A}^{X,+} _{a,j_X; a',j'_X} \) \frac{9}{8} \sum_{\substack{a.j_X, a', j'_X \\ (a,j_X)\neq (a',j_X')}} p_0(a) p^X_{\rm can}(j_X) P_{a,j_X  \to a',j'_X }\ln \frac{p_0(a)p^X_{\rm can}(j_X)}{p_0(a')p^X_{\rm can}(j'_X)} \nt \\
=&\( \sum_{\substack{a.j_X, a', j'_X \\ (a,j_X)\neq (a',j_X')}}\Di E_{a'}^2 {A}^{X} _{a,j_X; a',j'_X} \) \frac{9}{8} \sum_{\substack{a.j_X, a', j'_X \\ (a,j_X)\neq (a',j_X')}} p_0(a) p^X_{\rm can}(j_X) P_{a,j_X  \to a',j'_X }\ln \frac{p_0(a)p^X_{\rm can}(j_X)}{p_0(a')p^X_{\rm can}(j'_X)} \nt \\
=&\Theta_X \dot{\sigma}_X,
}
\end{widetext}
where we defined 
\eq{
\Theta_X:=\sum_{\substack{a.j_X, a', j'_X \\ (a,j_X)\neq (a',j_X')}} (\Di E_{a'})^2 A^X _{a,j_X; a',j'_X}.
}
In the second and seventh lines, we used  \eref{escape-trs}.
In the fifth line, we used the Schwarz inequality.
In the sixth line, we used an inequality $a\ln a/b+b-a \geq 8(a-b)^2/9(a+b)$ and the time-reversal symmetry of unitary evolution \eqref{trs}.

By applying the Schwarz inequality again, \eref{Markovkey} directly implies the desired inequality \eqref{Markovmain}:
\balign{
\sum_X|J_Q^X|\leq& \sum_X\sqrt{\Theta_X \dot{\sigma}_X} \nt \\
\leq& \sqrt{ \( \sum_X \Theta_X\) \( \sum_X \dot{\sigma}_X\) } \nt \\
=&\sqrt{\Theta (0)\dot{\sigma}}.
}


\begin{thebibliography}{99}





\bibitem{Carnot}
S. Carnot, {\it Reflections on the Motive Power of Fire and on Machines Fitted to Develop that Power}, Paris: Bachelier (1824).
\bib{JT}
J. P. Joule and W. Thomson, {\em On the Thermal Effects of Fluids in Motion. Part II}, Philosophical Transactions of the Royal Society of London, {\bf 144}, 321 (1854).
\bibitem{callen}
H. B. Callen, {\it Thermodynamics and an Introduction to Thermostatics}, 2nd ed. (John Wiley \& Sons, New York,
1985).

\bib{Lyeo04}
H. L. Lyeo, A. A. Khajetoorians, L. Shi, K. Pipe, R. J. Ram, A. Shakouri, and C. K. Shih, {\em Profiling the Thermoelectric Power of Semiconductor Junctions with Nanometer Resolution}, Science {\bf 303}, 816 (2004).
\bib{Casati07}
G. Casati, C. Mej\'{i}a-Monasterio, and T. Prosen, {\em Magnetically Induced Thermal Rectification}, Phys. Rev. Lett. {\bf 98}, 104302 (2007).
\bib{Taylor15}
E. Taylor and D. Segal, {\em Quantum Bounds on Heat Transport Through Nanojunctions}, Phys. Rev. Lett. {\bf 114}, 220401 (2015).

\bibitem{mahan}
G. D. Mahan, J.O. Sofo, {\em The best thermoelectric}. Proc. Natl. Acad. Sci. USA {\bf 93}, 7436. (1996).
\bibitem{mahan-rev}
G. D. Mahan, B. Sales, and J. Sharp, {\em Thermoelectric materials: New approaches to an old problem}, Phys. Today {\bf 50}, 42 (1997).
\bibitem{linke}
T.E. Humphrey, R. Newbury, R.P. Taylor, H. Linke, {\em Reversible Quantum Brownian Heat Engines for Electrons}. Phys. Rev. Lett. {\bf 89}, 116801 (2002).
\bibitem{majundar-rev}
A. Majumdar, {\em Thermoelectricity in Semiconductor Nanostructures}, Science {\bf 303}, 777 (2004).
\bibitem{dresselhaus}
M.S. Dresselhaus, G. Chen, M.Y. Tang, R.G. Yang, H. Lee, D.Z. Wang, Z.F. Ren, J.-P. Fleurial, and P. Gogna, {\em New Directions for Low-Dimensional Thermoelectric Materials}, Adv. Mater. {\bf 19}, 1043 (2007).
\bibitem{snyder}
G.J. Snyder and E.R. Toberer, {\em Complex thermoelectric materials}, Nature Materials {\bf 7}, 105 (2008).
\bib{Casati08}
G. Casati, C. Mej\'{i}a-Monasterio, and T. Prosen, {\em Incresing Thermoelectric Efficiency: A Dynamic Systems Approach}, Phys. Rev. Lett. {\bf 101}, 016601 (2008).
\bibitem{auto}
N. Shiraishi, {\em Attainability of Carnot efficiency with autonomous engines}, Phys. Rev. E {\bf 92}, 050101 (2015).
\bib{TH15}
H. Tajima and M. Hayashi, {\em Finite-size effect on optimal efficiency of heat engines}. Phys. Rev. E {\bf 96}, 012128 (2017).



\bibitem{Sekimoto-Sasa}
K. Sekimoto and S.-i. Sasa, {\em Complementarity relation for irreversible process derived from stochastic energetics}, J. Phys. Soc. Jpn. {\bf 66}, 3326 (1997).
\bibitem{Aurell}
E. Aurell, K. Gaw\c{e}dzki , C. Mej\'{\i}a-Monasterio, R. Mohayaee, P. Muratore-Ginanneschi, {\em Refined second law of thermodynamics for fast random processes}, J. Stat. Phys. {\bf 147}, 487 (2012).
\bib{Whitney}
R. S. Whitney. {\em Most Efficient Quantum Thermoelectric at Finite Power Output}, Phys. Rev. Lett. {\bf 112}, 130601 (2014).
\bibitem{Raz16}
O. Raz, Y. Suba\c{s}\i, and R. Pugatch, {\em Geometric Heat Engines Featuring Power that Grows with Efficiency}, Phys. Rev. Lett. {\bf 116}, 160601 (2016).














\bibitem{Benenti}
G. Benenti, K. Saito, and G. Casati, {\em Thermodynamic bounds on efficiency for systems with broken time-reversal symmetry}, Phys. Rev. Lett. {\bf 106}, 230602 (2011).

\bib{endo}
In the framework of endoreversible thermodynamics, the engine is assumed to be always in equilibrium, and the heat flux between the engine and the bath is assumed to be given by the Fourier law.

\bibitem{Sothmann}
B. Sothmann and M. B\"{u}ttiker, {\em Magnon-driven quantum-dot heat engine}, Europhys. Lett. {\bf 99}, 27001 (2012).
\bibitem{Brandner}
K. Brandner, K. Saito, and U. Seifert, {\em Strong bounds on Onsager coefficients and efficiency for three-terminal thermoelectric transport in a magnetic field}, Phys. Rev. Lett. {\bf 110}, 070603 (2013).
\bibitem{Brandner-full}
K. Brandner and U. Seifert, {\em Multi-terminal thermoelectric transport in a magnetic field: bounds on Onsager coefficients and efficiency}, New J. Phys. {\bf 15}, 105003 (2013).
\bibitem{Balachandran}
V. Balachandran, G. Benenti, and G. Casati, {\em Efficiency of three-terminal thermoelectric transport under broken time-reversal symmetry}, Phys. Rev. B {\bf 87}, 165419 (2013).

\bibitem{Brandner-new}
K. Brandner and U. Seifert, {\em Bound on thermoelectric power in a magnetic field within linear response}, Phys. Rev. E {\bf 91}, 012121 (2015).

\bib{Yamamoto}
K. Yamamoto, O. Entin-Wohlman, A. Aharony, and N. Hatano, {\em Efficiency bounds on thermoelectric transport in magnetic fields: The role of inelastic processes}, Phys. Rev. B {\bf 94}, 121402 (2016).



\bibitem{underdamped}
K. Brandner, K. Saito, and U. Seifert, {\em Thermodynamics of micro-and nano-systems driven by periodic temperature variations}, Phys. Rev. X {\bf 5}, 031019 (2015).
\bib{Bauer}
M. Bauer, K. Brandner, and U. Seifert, {\em Optimal performance of periodically driven, stochastic heat engines under limited control}, Phys. Rev. E {\bf 93}, 042112 (2016).
\bibitem{Proesmans}
K. Proesmans and C. Van den Broeck, {\em Onsager coefficients in periodically driven systems}, Phys. Rev. Lett. {\bf 115}, 090601 (2015).
\bibitem{Proesmans2}
K. Proesmans, B. Cleuren, and C. Van den Broeck, {\em Linear stochastic thermodynamics for periodically driven systems}, J. Stat. Mech. P023202 (2016).



\bibitem{Alla}
A. E. Allahverdyan, K. V. Hovhannisyan, A. V. Melkikh, and S. G. Gevorkian, {\em Carnot cycle at finite power: Attainability of maximal efficiency}, Phys. Rev. Lett. {\bf 111}, 050601 (2013).
\bib{EP}
M. Esposito and J. M. R. Parrondo, {\em Stochastic thermodynamics of hidden pumps}, Phys. Rev. E {\bf 91}, 052114 (2015)
\bibitem{Campisi}
M. Campisi and R. Fazio, {\em The power of a critical heat engine},Nature Commun. {\bf 7}, 11895 (2016).
\bibitem{Ponmurugan}
M. Ponmurugan, {\em Attainability of maximum work and the reversible efficiency from minimally nonlinear irreversible heat engines}, arXiv:1604.01912 (2016).
\bib{PE}
M. Polettini and M. Esposito, {\rm Carnot efficiency at divergent power output}, Europhys. Lett. {\bf 118}, 40003 (2017).


\bib{remark}
We remark that some of these researches utilize slightly different terminology from conventional one on such as ``finite power".


\bibitem{SS}
N. Shiraishi and T. Sagawa, {\em Fluctuation theorem for partially masked nonequilibrium dynamics},Phys. Rev. E {\bf 91}, 012130 (2015).
\bibitem{SIKS}
N. Shiraishi, S. Ito, K. Kawaguchi, and T. Sagawa, {\em Role of measurement-feedback separation in autonomous Maxwell's demons }, New J. Phys. {\bf 17}, 045012 (2015).
\bibitem{SMS}
N. Shiraishi, T. Matsumoto, and T. Sagawa, {\em Measurement-feedback formalism meets information reservoirs}, New J. Phys. {\bf 18}, 013044 (2016).

\bib{SST}
N. Shiraishi, K. Saito, and H. Tasaki, {\em Universal Trade-Off Relation between Power and Efficiency for Heat Engines}, Phys. Rev. Lett. {\bf 117}, 190601 (2016).


\bibitem{steeneken2010}
P. G. Steeneken, K. Le Phan, M. J. Goossens, G. E. J. Koops, G. J. A. M. Brom, C. van der Avoort and J. T. M. van Beek, {\em Piezoresistive heat engine and refrigerator}, Nat. Phys. {\bf 7}, 354 (2011).
\bibitem{crivellari2014}
M. Ribezzi-Crivellari and F. Ritort, {\em Free-energy inference from partial work measurements in small systems}, Proc. Natl. Acad. Sci. USA {\bf 111}, E3386 (2014).
\bibitem{koski2014}
J .V. Koski, V. F. Maisi, T. Sagawa, and J. P. Pekola, {\em Experimental Observation of the Role of Mutual Information in the Nonequilibrium Dynamics of a Maxwell Demon}, Phys. Rev. Lett. {\bf 113}, 030601 (2014).
\bibitem{rosnagel2015}
J. Rosnagel, S.T. Dawkins, K. N. Tolazzi, O. Abah, E. Lutz, F. S. Kaler and Kilian Singer, {\em A single-atom heat engine},  Science {\bf 352},  325 (2016).



\bib{LR}
E. Lieb and D. Robinson, {\em The finite group velocity of quantumspin systems}, Commun. Math. Phys. {\bf 28}, 251 (1972).
\bib{Has04}
M. B. Hastings, {\em Lieb-Schultz-Mattis in higher dimensions}, Phys. Rev. B {\bf 69}, 104431 (2004).
\bib{NOS06}
B. Nachtergaele, Y. Ogata, and R. Sims, {\em Propagation of correlations in quantum lattice systems}, J. Stat. Phys. {\bf 124}, 1 (2006).
\bib{HK}
M. B. Hastings and T. Koma, {\em Spectral gap and exponential decay of correlations}, Commun. Math. Phys. {\bf 265}, 781 (2006).
\bib{CSE08}
M. Cramer, A. Serafini, and J. Eisert, {\em Locality of dynamics in general harmonic quantum systems}, arXiv:0803.0890 (2008).
\bib{Schuch11}
N. Schuch, S. K. Harrison, T. J. Osborne, and J. Eisert, {\em Information propagation for interacting-particle systems}, Phys. Rev. A {\bf 84}, 032309 (2011).



\bib{NS06}
B. Nachtergaele and R. Sims, {\em Lieb-Robinson bounds and the exponential clustering theorem}, Commun. Math. Phys. {\bf 265}, 119 (2006).
\bib{Has07}
M. B. Hastings, {\em An area law for one-dimensional quantum systems}, J. Stat. Mech. {\bf 09}, P08024 (2007).

\bib{HM09}
M. B. Hastings and S. Michalakis, {\em Quantization of Hall conductance for interacting electrons on a torus}, Commun. Math. Phys. {\bf 334}, 433 (2015).

\bib{IKS}
E. Iyoda, K. Kaneko, and T. Sagawa, {\em Emergence of the information-thermodynamics link and the fluctuation theorem for pure quantum states}, arXiv:1603.07857 (2016).
\bib{KMS}
T. Kuwahara, T. Mori, and K. Saito, {\em Floquet–Magnus theory and generic transient dynamics in periodically driven many-body quantum systems}, Ann. Phys. {\bf 367}, 96 (2016)


\bibitem{AN}
S. Amari and H. Nagaoka, {\it{Methods of Information Geometry}}. (Oxford University Press, 2000).
\bibitem{secondused}
H. Tajima and M. Hayashi, in preparation




\bib{CC}
P. Calabrese and J. Cardy, {\em Time Dependence of Correlation Functions Following a Quantum Quench}, Phys. Rev. Lett. {\bf 96}, 136801 (2006).
\bib{CDEO}
M. Cramer, C. M. Dawson, J. Eisert, and T. J. Osborne, {\em Exact Relaxation in a Class of Nonequilibrium Quantum Lattice Systems}, Phys. Rev. Lett. {\bf 100}, 030602 (2008).
\bib{ISPU}
T. N. Ikeda, N. Sakumichi, A. Polkovnikov, and M. Ueda, {\em The second law of thermodynamics under unitary evolution and external operations}, Ann. Phys. {\bf 354}, 338 (2015).
\bib{MAMW}
M. P. Mueller, E. Adlam, L. Masanes, N. Wiebe, {\em Thermalization and Canonical Typicality in Translation-Invariant Quantum Lattice Systems}, Commun. Math. Phys. {\bf 340}, 499 (2015).

\bib{BDZ}
I. Bloch, J. Dalibard, and W. Zwerger, {\em Many-body physics with ultracold gases}, Rev. Mod. Phys. {\bf 80}, 885 (2008).
\bib{Tro}
S. Trotzky,	Y-A. Chen, A. Flesch, I. P. McCulloch, U. Schollwöck,	J. Eisert, and I. Bloch, {\em Probing the relaxation towards equilibrium in an isolated strongly correlated one-dimensional Bose gas}, Nat. Phys. {\bf 8}, 325 (2012).
\bib{Kau}
A. M. Kaufman, M. E. Tai, A. Lukin, M. Rispoli, R. Schittko, P. M. Preiss, and M. Greiner, {\em Quantum thermalization through entanglement in an isolated many-body system}, Science {\bf 353}, 794 (2016).



\bib{comment-driven}
Therefore, the laser cooling, for example, is out of our scope.


\bib{KSH}
K. Kanazawa, T. Sagawa, and H. Hayakawa, {\em Heat conduction induced by non-Gaussian athermal fluctuations}, Phys. Rev. E {\bf 87}, 052124 (2013).
\bib{DBS}
A. Dechant, A. Baule, and S.-i. Sasa, {\em Gaussian white noise as a resource for work extraction}, Phys. Rev. E {\bf 95}, 032132 (2017).


\bib{QFT}
H. Tasaki, {\em Jarzynski Relations for Quantum Systems and Some Applications}, arXiv:cond-mat/0009244 (2000).
\bib{CHT}
M. Campisi, P. H\"{a}nggi, and P. Talkner, {\em Colloquium: Quantum fluctuation relations: Foundations and applications}, Rev. Mod. Phys. {\bf 83}, 771 (2011).
\bib{Brandao}
F. Brand\~{a}o, M. Horodecki, N. Ng, J Oppenheim, and S. Wehner, {\em The second laws of quantum thermodynamics}, Proc. Nat. Ac. Sci. {\bf 112}, 3275 (2014).

\bib{req-cycle}
This requirement adopts the standpoint that the baths are removed at the end of the cyclic process, and we refresh the baths if we perform the cyclic process twice and more.

\bib{HO}
M. Horodecki and J. Oppenheim, {\em Fundamental limitations for quantum and nanoscale thermodynamics}, Nat. Comm. {\bf 4}, 2059 (2013).
\bib{SSP}
P. Skrzypczyk, A. J. Short, and S. Popescu, {\em Work extraction and thermodynamics for individual quantum systems}, Nat. Comm. {\bf 5}, 4185 (2014).
\bib{TWO}
H. Tajima, E. Wakakuwa, and T. Ogawa, {\em Large Deviation implies First and Second Laws of Thermodynamics}, arXiv:1611.06614 (2016).

\bib{oneshot1}O. C. O. Dahlsten, R. Renner, E. Rieper, and V. Vedral, New. J. Phys.\textbf{13}, {\em Inadequacy of von Neumann entropy for characterizing extractable work} 053015, (2011).  
\bib{oneshot2}L. Rio, J. Aberg, R. Renner, O. Dahlsten, and V. Vedral, {\em The thermodynamic meaning of negative entropy}, Nature \textbf{474}, 61, (2011). 
\bib{oneshot3}J. Aberg, {\em Truly work-like work extraction via a single-shot analysis}, Nat. Commun. \textbf{4}, 1925 (2013).
\bib{Car2}S. Popescu, {\em Maximally efficient quantum thermal machines: The basic principles}, arXiv:1009.2536.(2010).
\bib{Popescu2015}Y. Guryanova,	S. Popescu,	A. J. Short,	R. Silva and P. Skrzypczyk, {\em Thermodynamics of quantum systems with multiple conserved quantities}, Nat. Comm. \textbf{7}, 12049, (2016).
\bib{review}G. Gour, M. P. Muller, V. Narasimhachar, R. W. Spekkens, N. Y. Halpern, {\em The resource theory of informational nonequilibrium in thermodynamics}, Phys. Rep. \textbf{583}, 1 (2015).
\bib{catalyst}J. \r{A}berg, {\em Catalystic coherence} Phys. Rev. Lett. \textbf{113}, 150402 (2014).
\bib{Malabarba}A. S. L. Malabarba, A. J. Short, P. Kammerlander, {\em Clock-driven quantum thermal engines}, New. J. Phys. \textbf{17}, 045027 (2015).
\bib{m-based}
M. Hayashi and H. Tajima, {\em Measurement-based Formulation of Quantum Heat Engine}, Phys. Rev. A 95, 032132 (2017).
\bib{T-W}
H. Tajima and E. Wakakuwa, {\em Regularized Boltzmann entropy determines macroscopic adiabatic accessibility},  arXiv:1601.00487, (2016).
\bib{MTH}Y. Morikuni, H. Tajima, N. Hatano, {\em Quantum Jarzynski equality of measurement-based work extraction}, Phys. Rev. E 95, 032147 (2017).
\bib{Tasaki2015}H. Tasaki, {\em Quantum statistical mechanical derivation of the second law of thermodynamics: a hybrid setting approach}, Phys. Rev. Lett. 116, 170402 (2016).




\bib{NC}
M. A. Nielsen and I. L. Chuang, {\it Quantum Computation and Quantum Information}, Cambridge University Press (2000).
\bib{Jay}
E. T. Jaynes, {\em Information Theory and Statistical Mechanics. II}, Phys. Rev. {\bf 108}, 171 (1957).

\bib{def-norm}
The operator norm of $A$ is defined as $\| A\| :=\max_{\ket{\psi}} \| A\ket{\psi} \| /\| \ket{\psi} \|$.
The norm of the state $\ket{\psi}$ is given by using the conventional inner product: $\| \ket{\psi}\| :=\sqrt{\braket{\psi |\psi}}$.

\bib{def-c}
The constant $c$ is defined as a quantity satisfies the following relation:
\eq{
\sum_z e^{-\mu d(x,z)+d(z,y)}\leq c e^{-\mu d(x,y)}, \nt
}
where $z$ runs all site such that there exist two local Hamiltonians whose supports respectively contain $\{ x,z\}$ and $\{ y,z\}$.


\bib{Cham}
P. Chambadal, {\em Les centrales nucl{\'e}aires}, (Colin 1957).
\bib{Nov}
I. Novikov, {\em The efficiency of atomic power stations (a review)}, J. Nuc. Energy {\bf 7}, 125 (1958).
\bib{CA}
F. L. Curzon and B. Ahlborn, {\em Efficiency of a Carnot engine at maximum power output}, Am. J. Phys. {\bf 43}, 22 (1975).

\bibitem{Broeck}
C. Van den Broeck, {\em Thermodynamic efficiency at maximum power}, Phys. Rev. Lett. {\bf 95}, 190602 (2005).
\bib{ELB}
M. Esposito, K. Lindenberg, and C. Van den Broeck, {\em Universality of efficiency at maximum power}, Phys. Rev. Lett. {\bf 102}, 130602 (2009).
\bib{Esposito}
M. Esposito, R. Kawai, K. Lindenberg, and C. Van den Broeck, {\em Efficiency at Maximum Power of Low-Dissipation Carnot Engines}, Phys. Rev. Lett. {\bf 105}, 150603 (2010).

\bib{Taka}
M. Takahashi, {\it Thermodynamics of One-Dimensional Solvable Models}, Cambridge University Press (1999).

\bib{Sagawa}
T. Sagawa, {\it Lectures on Quantum Computing, Thermodynamics and Statistical Physics}, 8, 127 (2012).

\bibitem{breuer} 
H. Breuer and F. Petruccione, {\it Theory of Open Quantum Systems}, (Oxford, Oxford, 2002).

\bib{SSarxiv}
N. Shiraishi and K. Saito, {\em Incompatibility between Carnot efficiency and finite power in Markovian dynamics}, arXiv:1602.03645 (2016, unpublished).

\end{thebibliography}
\end{document}